\documentclass[12pt]{article}
\usepackage{amsmath}
\usepackage{amssymb}
\usepackage{graphicx,psfrag,epsf}
\usepackage{enumerate}
\usepackage{natbib}
\usepackage{bm}
\usepackage{epstopdf}
\usepackage{comment}
\usepackage{color}
\usepackage{algorithm2e}
\usepackage{float}
\usepackage{url} 

\newcommand{\blind}{1}

\def \pr{\mbox{pr}}
\def \diag{\mbox{diag}}
\def \var{\mbox{var}}
\newcommand{\pkg}[1]{{\fontseries{b}\selectfont #1}}

\def\bbeta{\bm \beta}
\def\btheta{\bm \theta}

\def\boldeta{\bm \eta}

\def \ep{\epsilon}
\def \mmin{\wedge}
\def \eigen{\mbox{eigen}}

\newtheorem{theorem}{Theorem}

\newtheorem{lemma}{Lemma}

\newtheorem{remark}{Remark}[section]

\begin{document}

\def\spacingset#1{\renewcommand{\baselinestretch}%
{#1}\small\normalsize} \spacingset{1}


\if1\blind
{
 \title{\bf {Scalable Algorithms for Large Competing Risks Data }}
  \author{Eric S. Kawaguchi$^1$ \\
  Jenny I. Shen$^2$ \\
  Marc A. Suchard$^{3, 4, 5}$ \\
  and
  Gang Li$^{3,4,*}$  \\
  $^1$  Division of Biostatistics and Epidemiology, University of Southern California \\
  $^2$ Division of Nephrology and Hypertension \\ Los Angeles Biomedical Institute at Harbor-UCLA Medical Center \\
  $^3$ Department of Biostatistics, University of California, Los Angeles  \\
  $^4$ Department of Biomathematics, University of California, Los Angeles  \\
  $^5$ Department of Human Genetics,  University of California Los Angeles \\
    $^*$email: vli@ucla.edu}
    \date{}
  \maketitle
} \fi

\if0\blind
{
  \bigskip
  \bigskip
  \bigskip
  \begin{center}
    {\LARGE\bf Scalable Algorithms for Large Competing Risks Data}
\end{center}
  \medskip
} \fi

\bigskip
\begin{abstract}
This paper develops two orthogonal contributions to scalable sparse regression for  competing risks time-to-event data. First, we study and accelerate the broken adaptive ridge method (BAR), an $\ell_0$-based  iteratively reweighted $\ell_2$-penalization algorithm that achieves sparsity in its limit, in the context of the Fine-Gray (1999) proportional subdistributional hazards (PSH) model. In particular, we derive a new algorithm for BAR regression, named cycBAR, that  performs cyclic  update of each coordinate using an explicit thresholding formula. The new cycBAR algorithm effectively avoids fitting multiple reweighted $\ell_2$-penalizations and thus yields  impressive speedups over the original BAR algorithm.  Second, we address a pivotal computational issue related to fitting the PSH model. Specifically, the computation costs of the log-pseudo likelihood and its derivatives for PSH model grow at the rate of $O(n^2)$ with the sample size $n$ in current implementations. We propose a novel forward-backward scan algorithm that reduces the computation costs  to $O(n)$. The proposed method  applies to both unpenalized and penalized estimation for the PSH model and  has exhibited drastic speedups over current implementations. Finally, combining the two  algorithms can yields $>1,000$ fold speedups  over the original BAR algorithm. Illustrations of the impressive scalability of our proposed algorithm for large competing risks data are given using both simulations and a United States Renal Data System data. 
\end{abstract}
\noindent%
\vfill

\spacingset{1.5} 
\section{Introduction}
Advancing informatics tools make large-scale data such as electronic health record (EHR) data and genomic data routinely
accessible to researchers. 
This data deluge offers unprecedented opportunities for new and innovative approaches to improve research and learning  \citep{schuemie2017honest}. However, it also presents new computational challenges and barriers for quantitative researchers as many current statistical methodologies and computational tools may grind to a halt as the sample size ($n$) and/or number of covariates ($p_n$) grow large. 
Such challenges are particularly common in time-to-event data analysis where the likelihood function (such as the partial likelihood for the Cox model with  data) and its derivatives typically require $O(n^2)$ number of operations, which will explode quickly as $n$ increases. The computational burden can be further aggravated as  the number of covariates ($p_n$) increases.
Statistical methods coupled with high-performance algorithms are critically needed for large-scale time-to-event data analysis.

This paper aims to develop high-performance computational methods for  large-scale competing risks time-to-event data analysis by addressing two orthogonal computational challenges due to  large $p_n$ and large $n$ respectively.
First, we develop a scalable $\ell_0$-based method for simultaneous variable selection and parameter estimation for the large $p_n$ problem.
It is well known that $\ell_0$-penalized regression is natural for variable selection, but is computationally NP hard and not scalable to even moderate $p_n$. 
As a scalable approximation to $\ell_0$-penalized regression, the broken adaptive ridge \textsc{BAR} estimator, defined as the limit of an $\ell_0$-based  iteratively reweighted
$\ell_2$-penalization algorithm, has been recently studied for simultaneous variable selection
and parameter estimation and shown to possess some desirable selection, estimation, and grouping properties under various model settings
(see, e.g., \cite{zhao2018variable},  \cite{dai2018broken}, \cite{zhao2019simultaneous}, and \cite{zhao2019simultaneous2}).
However, while
previous research has focused on studying the statistical properties of BAR methodology,
 the feasibility of applying BAR to large-scale data has yet to be explored. To this end, we note that
BAR requires fitting multiple reweighted $\ell_2$-penalized regressions until convergence, which can potentially create a computational bottleneck if a large number of iterations is needed for convergence.  As demonstrated in Table 1 of Section \ref{s2:data}, BAR can grind to a halt for large scale  data.  Second, we address a pivotal computational issue specifically related to fitting the PSH model when $n$ is large.   As discussed later in Section \ref{s2:linear}, computation of the log-pseudo likelihood and its derivatives for the PSH model  involves $O(n^2)$ number of operations, which presents a critical computational barrier as $n$ becomes large. Moreover, because the computations involve weighted sums over some risk sets where the risks sets are not monotone over time and the weights are subject-specific,  commonly used efficient computational techniques for fitting the \cite{cox1972regression} model do not apply to the PSH model. To the best of
our knowledge, no algorithm exists in the literature that reduces the computational cost for the PSH model from $O(n^2)$ to a lower order.

In addressing the aforementioned computational challenges for large data,  the contribution of this paper is two folds: 
\begin{enumerate}

\item
We propose a novel cyclic coordinate-wise update algorithm for BAR, referred to as \textsc{cycBAR}, by deriving an explicit analytic coordinate-wise update for a fixed-point problem whose unique solution approximates the \textsc{BAR}  estimator. 
Because the \textsc{cycBAR} algorithm avoids carrying out  iteratively reweighted $\ell_2$-penalizations, it  can result in substantial gains in computational efficiency.  We emphasize that
the application of the \textsc{cycBAR} algorithm over the original BAR method is not limited to the PSH model and spans a variety of models and data settings such as generalized linear models and time-to-event models, as well as sparse signal reconstruction \citep{gorodnitsky1997sparse} and compressive sensing \citep{candes2008enhancing,chartrand2008iterative,gasso2009recovering,
	daubechies2010iteratively,wipf2010iterative} where the $\ell_0$-based iteratively reweighted $\ell_2$-penalization algorithm are popularly used.
In our numerical studies (Section \ref{s2:timing}, Figure 1(b)), \textsc{cycBAR} showed marked reduction in runtime 
over the standard BAR.

\item
By exploiting the special structure of the risk set and the subject-specific weight functions associated with the Fine-Gray pseudo likelihood and its derivatives, we derive a novel forward-backward scan algorithm to reduce their computational costs from $O(n^2)$ to $O(n)$,  allowing one to analyze competing risks data much quicker than current approaches.  
We have observed in empirical studies, e.g. Figure 1(c) in Section 3.3, that the forward-backward scan algorithm 
can yield dramatic speedups over standard implementations.  We point out that our proposed forward-backward scan algorithm for the PSH model is not specific to the BAR method and can be applied  to accelerate other penalized  regression methods such as  LASSO  \citep{tibshirani1996regression}, SCAD \citep{fan2001variable}, 
adaptive LASSO \citep{zou2006adaptive}, and MCP \citep{zhang2010nearly} for the PSH model \citep{fu2017penalized}, and the unpenalized estimation method of \cite{fine1999proportional}, as well as to hypothesis testing and cumulative incidence estimation   
for the PSH model.
\end{enumerate}

The rest of this article is organized as follows. In Section \ref{s2:estimator}, we review the mathematical formulation of competing risks data and the \cite{fine1999proportional}  proportional subdistribution hazards model. Section \ref{s2:pshbar}, introduces the BAR estimator for the PSH model and refers its asymptotic properties to the Online Supplementary Material. Section \ref{s2:cyclic} derives the cyclic coordinate-wise BAR algorithm. The forward-backward scan method for the PSH model is described in Section \ref{s2:linear}. Section \ref{s2:simulation} presents some simulation studies to demonstrate the computational efficiency gains of both the \textsc{cycBAR} and forward-backward scan algorithms. A proof-of-concept real data example for fitting large-scale competing risks data is provided in Section \ref{s2:data} using a
subset of the United States Renal Data System (USRDS). Concluding remarks are given in  Section \ref{s2:discussion}. 
The proposed method has been implemented in an \textsf{R} package, named \pkg{pshBAR}, which is available at \verb+https://github.com/erickawaguchi/pshBAR+.

\section{Methodology}
\label{s2:methods}

\subsection{Competing risks data, model, and parameter estimation}
\label{s2:estimator}
Competing risks time-to-event data arises frequently in clinical trials, reliability testing, social science, and many other fields \citep{prentice1978,  pintilie2006competing, putter2007}. Competing risks  occur when individuals are susceptible to more than one types of possibly correlated events or causes and the occurrence of one event precludes the others from happening. 
For example, one may wish to study time until first kidney transplant for kidney dialysis patients with end-stage renal disease. Then
terminating events such as death, renal function recovery, or discontinuation of dialysis are competing risks as their occurrence will prevent subjects from receiving a transplant. 
For $i = 1, \ldots, n$, let $T_i$, $C_i$, $\ep_i$, and $\mathbf{z}_i$ be the event time, possible right-censoring time,  cause (event type), and a $p_n$-dimensional vector of time-independent covariates, respectively, for subject $i$. Without loss of generality assume there are two event types $\ep \in \{1, 2\}$ where $\ep = 1$ is the event of interest and $\ep = 2$ is the competing risk. With the presence of right censoring, we generally observe $X_i = T_i \mmin C_i$, $\delta_i = I(T_i \leq C_i)$, where $a \mmin b = \min(a, b)$ and $I(\cdot)$ is the indicator function. 
Competing risks data consists of $n$ independent and identically distributed quadruplets $\{(X_i, \delta_i, \delta_i \ep_i, \mathbf{z}_i\}_{i=1}^n$. Assume that there exists a $\tau$ such that (1) for some arbitrary time $t$, $t \in [0, \tau]$; (2) $\Pr(T_i > \tau) > 0$ and $\Pr(C_i > \tau) >0$  for all $i = 1,\ldots, n$.

An important quantity for competing risks data is the cumulative incidence function (CIF), which describes the probability of failing from a certain cause of interest before the other causes.
The CIF for cause 1 events conditional on the covariates is defined as $F_1(t; \mathbf{z}) = \Pr(T \leq t, \epsilon = 1|\mathbf{z})$. To model  $F_1(t; \mathbf{z})$, \cite{fine1999proportional} introduced the now popular
proportional subdistribution hazards (PSH) model: 
\begin{align}
\label{eq2:pshmodel}
h_1(t| \mathbf{z}) = h_{10}(t) \exp(\mathbf{z}^\prime \bbeta),
\end{align}
where \begin{align*}
h_1(t| \mathbf{z}) \! =\!  \lim_{\Delta t \to 0} \frac{\Pr\{t \leq T \leq t + \Delta t, \epsilon = 1 | T \geq t \cup (T \leq t \cap \epsilon \neq 1), \mathbf{z}\}}{\Delta t} 
\! = -\frac{d}{dt} \log\{1 - F_1(t; \mathbf{z})\}
\end{align*}
is a subdistribution hazard \citep{gray1988class}, 
$h_{10}(t)$ is a completely unspecified baseline subdistribution hazard, and $\bbeta$ is a $p_n \times 1$ vector of regression coefficients. 
As \cite{fine1999proportional} mentioned, the risk set associated with $h_1(t; \mathbf{z})$ is somewhat 
counterfactual as it includes subjects who are still at risk $(T \geq t)$ and those who have already observed the competing risk prior to time $t$ ($T \leq t \cap \epsilon \neq 1$). However, this construction is useful  for direct modeling of the CIF.

Inference for the PSH model based on the following log-pseudo likelihood \citep{fine1999proportional}:
\begin{align}
\label{eq2:psh_pll}
l(\bbeta) = \sum_{i=1}^n \int_0^\tau \left( \mathbf{z}_i^\prime \bbeta - \log \left\{ \sum_j \hat{w}_j(s)Y_j(s)\exp(\mathbf{z}_j^\prime \bbeta)\right\}\right) \times \hat{w}_i(s) dN_i(s),
\end{align}
where $N_i(t) = I(T_i \leq t, \ep_i = 1)$, $Y_i(t) = 1 - N_i(t-)$, $\hat{w}_i(t)$ is a time-dependent weight  for subject $i$ at time $t$ defined as $\hat{w}_i(t) = I(C_i \geq T_i \mmin t)\hat{G}(t)/\hat{G}(X_i \mmin t)$, and $\hat{G}(t)$ is the \cite{kaplan1958nonparameteric} estimate for  $G(t) = \Pr(C \geq t)$,  the survival function of the censoring variable $C$. Note that, for any subject $i$ and time $t$, $\hat{w}_i(t)Y_i(t) = 0$ if an individual is right censored or has experienced the event of interest; and  $\hat{w}_i(t)Y_i(t) = 1$ if $t < X_i$, and $\hat{w}_i(t)Y_i(t) = \hat{G}(t) / \hat{G}(X_i)$ for events due to the competing risk. 

Commonly-used optimization routines to estimate the parameters of the PSH model typically require the calculation of the log-pseudo likelihood (\ref{eq2:psh_pll}), the score function 
\begin{align}
\label{eq2:score}
\dot{l}_j(\bbeta) = & \sum_{i=1}^n I(\delta_i\ep_i = 1) z_{ij} - \sum_{i = 1}^n I(\delta_i\ep_i = 1) \frac{ \sum_{k \in R_i} z_{kj} \tilde{w}_{ik}  \exp(\eta_k)}{\sum_{k \in R_i} \tilde{w}_{ik} \exp(\eta_k)},
\end{align}
and, in some cases, the Hessian diagonals
\begin{align}
\label{eq2:hess}
\ddot{l}_{jj}(\bbeta) = & \sum_{i=1}^n I(\delta_i\ep_i = 1) \left[ \frac{ \sum_{k \in R_i} z_{kj}^2 \tilde{w}_{ik}  \exp(\eta_k)}{\sum_{k \in R_i} \tilde{w}_{ik} \exp(\eta_k)} -  \left\{ \frac{ \sum_{k \in R_i} z_{kj} \tilde{w}_{ik}  \exp(\eta_k)}{\sum_{k \in R_i} \tilde{w}_{ik} \exp(\eta_k)} \right\}^2 \right], 
\end{align}
where  $$\tilde{w}_{ik} = \hat{w}_k(X_i) = \hat{G}(X_i) / \hat{G}(X_i \mmin X_k), \quad k \in R_i,$$ 
$R_i = \{y:(X_y \geq X_i) \cup (X_y \leq X_i \cap \epsilon_y =  2)\}$ and
$\eta_k = \mathbf{z}_k^\prime\bbeta$.
Direct calculations using the above formulas will need $O(n^2)$ operations due to the the double summations and is computationally taxing for large $n$. We will show how to calculate the double summation linearly in Section \ref{s2:linear}, allowing us to calculate these quantities in $O(n)$ time.


\subsection{Broken adaptive ridge estimation for the proportional subdistribution hazards model}
\label{s2:pshbar}
Penalized regression is useful for simultaneous variable selection and parameter estimation and has recently been introduced to the PSH model for competing risks data \citep{ha2014variable, fu2017penalized, ahn2018group, hou2018high}.  Below we extend the broken adaptive ridge (BAR) estimator to the PSH model. 

Let $l(\bbeta)$ be the log-pseudo likelihood defined by (\ref{eq2:psh_pll}).
The BAR estimator of $\bbeta$ starts with an initial $\ell_2$-penalized (or ridge) estimator
\begin{equation}
\label{eq2:init_ridge}
\hat{\bbeta}^{(0)} = \mbox{arg}\min_{\bbeta} \{-2l(\bbeta) + \xi_n \sum_{j=1}^{p} \beta_j^2\},
\end{equation}
which is updated iteratively by a reweighted $\ell_2$-penalized estimator
\begin{equation}
\label{eq2:l0_approx}
\hat{\bbeta}^{(k)}  = \mbox{arg} \min_{\bbeta}  \left\{-2 l(\bbeta) + \lambda_n \sum_{j =1}^{p} \frac{\beta_j^2}{|\hat\beta_j^{(k-1)}|^2} \right\}, \quad k\ge 1,
\end{equation}
where  $\xi_n$ and $\lambda_n$ are non-negative penalization tuning parameters.
The BAR estimator of $\bbeta$ is defined as the limit of this iterative algorithm: 
\begin{equation}
\label{eq2:bar_est}
\hat{\bbeta} = \lim_{k \to \infty} \hat{\bbeta}^{(k)},
\end{equation}
which can be viewed as a surrogate  to $\ell_0$-penalized regression.

Note that adaptively reweighting the penalty of a coefficient by the inverse of its squared estimate from the previous iteration allows each coefficient to be penalized differently. At each successive iteration, coefficients whose true values are zero will have larger penalties that will shrink the estimate further towards zero. We have shown in Section S1 of the Online Supplementary Material 
that the BAR estimator has an oracle property for selection and estimation and a grouping properties for highly correlated covariates.

The BAR estimator can be implemented using the algorithm outlined in Section S2.1 Algorithm S1 of the Online Supplementary Material in which
cyclic coordinate decent (CCD) algorithm is employed for each reweighted $\ell_2$-penalized regression. 
Because the algorithm runs a sequence ($k = 0, 1, \ldots$) of adaptively reweighted ridge regressions, it adds an extra layer
of computational complexity as compared to other popular single-step penalization methods such as LASSO and 
can create a bottleneck when a large number of iterations is needed.  Moreover,  because 
ridge regression is not sparse and thus the limit
is never achieved at any given step of the BAR algorithm, an arbitrarily small cutoff value $\epsilon^*$ has to be used to induce sparsity in Algorithm S1 (line 18), which is an unpleasant feature. Below we show that these issues can be avoided using 
a new cyclic BAR algorithm.

\subsection{A cyclic coordinate-wise BAR algorithm}
\label{s2:cyclic}
In this section, we derive a fast cyclic coordinate-wise BAR algorithm that will result in the elimination of performing multiple ridge regressions and
avoid using a cutoff $\epsilon^*$ to introduce sparsity as required by the original BAR algorithm (Algorithm S1 in the Online Supplementary Matieral).
For 
a consistent estimate $\tilde{\bbeta}$ of $\bbeta$, consider the Cholesky decomposition $-\ddot{l}(\tilde\bbeta) = \tilde{\mathbf{X}}^\prime\tilde{\mathbf{X}}$ and define $\tilde{\mathbf{y}} = (\tilde{\mathbf{X}}^\prime)^{-1}\{-\ddot{l}(\tilde\bbeta)\tilde\bbeta + \dot{l}(\tilde\bbeta)\}$ as the pseudo-response vector. Approximating the negative log-{pseudo} likelihood by $-l(\bbeta) \approx (1/2)(\tilde{\mathbf{y}} - \tilde{\mathbf{X}}\bbeta)^\prime(\tilde{\mathbf{y}} - \tilde{\mathbf{X}}\bbeta)$ using a second-order Taylor expansion in (\ref{eq2:l0_approx}) leads to the following solution
$$
\hat{\bbeta}^{(k)} = g(\hat{\bbeta}^{(k-1)}),
$$
where 
$
g(\bbeta) = \{\tilde{\mathbf{X}}^\prime\tilde{\mathbf{X}} + \lambda_nD(\bbeta)\}^{-1} \tilde{\mathbf{X}}^\prime\tilde{\mathbf{y}}.
$
and $D(\bbeta) = \diag(\beta_1^{-2}, \ldots, \beta_{p_n}^{-2})$. Hence, as $k\to\infty$, the limit of the sequence $\{ \hat{\bbeta}^{(k)}\}$ is 
the fixed point of the function $g(\cdot)$ or the solution of $g(\bbeta) =\bbeta$.

The next theorem shows that each component of the fixed-point solution of $g$  can be expressed
as a function of all other components. The proof is deferred to Section S1.5 of the Online Supplementary Material.

\begin{theorem}
\label{th2:fixed}
Let $\hat\bbeta$ be the fixed-point solution of $g(\cdot)$. Then, for each $j=1,\ldots, p_n$, the $j$th component of $\hat\bbeta$ can be expressed as follows 
\begin{align}
\label{eq2:fpoint_uni}
\hat{\beta}_j = g_j(\hat\bbeta_{-j})\equiv
 \begin{cases} 
      0, & \mbox{if }  | b_j| < 2\sqrt{\lambda_n \tilde{\mathbf{x}}_j^\prime\tilde{\mathbf{x}}_j}, \\
     \frac{ b_j + sign( b_j) \sqrt{ ( b_j)^2 - 4 \lambda_n \tilde{\mathbf{x}}_j^\prime\tilde{\mathbf{x}}_j
}}{2\tilde{\mathbf{x}}_j^\prime\tilde{\mathbf{x}}_j},
 & \mbox{otherwise,}
   \end{cases}
\end{align}
where 
$
 b_j =  \tilde{\mathbf{x}}_j^\prime(\tilde{\mathbf{y}} -  \sum_{i \neq j} \tilde{\mathbf{x}}_i \hat\beta_i) 
$ and
 $\hat\bbeta_{-j} =(\hat\beta_1,\ldots, \hat\beta_{j-1},\hat\beta_{j+1}, \dots, \hat\beta_{p_n})^\prime$ .
\end{theorem}

The above result motivates our cyclic coordinate-wise broken adaptive ridge (\textsc{cycBAR}) algorithm which
performs cyclic coordinate-wise updates for the fixed point of $g(\cdot)$ using equation (\ref{eq2:fpoint_uni}) as outlined in Algorithm \ref{alg2:cwbar} below. In Algorithm \ref{alg2:cwbar}, $\tilde{\mathbf{X}}$ and $\tilde{\mathbf{y}}$ are initially estimated using the initial ridge estimate ${\bbeta}^{(0)}$
and then subsequently updated at step $s$ using  the previous estimate ${\bbeta}^{(s-1)}$ for $s\ge 1$. 
Consequently, at step $s$,  we have
\begin{align*}
 b_j^{(s)} & \equiv 
 \tilde{\mathbf{x}}_j^\prime \left\{\tilde{\mathbf{y}} -  \sum_{i \neq j} \tilde{\mathbf{x}}_i \beta^{(s-1)}_i \right\} 
 = - \ddot{l}_{jj}(\bbeta^{(s-1)})\beta_j^{(s-1)} +\dot{l}_j(\bbeta^{(s-1)}) ,\quad \mbox{for $j=1,\ldots,p_n$,}
\end{align*}
where 
$\dot{l}_j(\bbeta) $
is the $j$th element of $ -\dot{l}(\bbeta) $ and $ -\ddot{l}_{jj}(\bbeta) $ 
is the $j$th diagonal element of $ \ddot{l}(\bbeta) $. 

\RestyleAlgo{boxruled}
\LinesNumbered
\begin{algorithm}[t]
  \footnotesize
   \caption{The \textsc{cycBAR} algorithm}
\label{alg2:cwbar}
\SetAlgoLined
 Set $\bbeta^{(0)} = \hat{\bbeta}_{ridge}$\;
 \For{$s= 1, 2, \ldots$}{
$\#$ Enter cyclic coordinate-wise BAR algorithm\\
  \For{$j = 1, \ldots p_n$}{
  Calculate 
  $c_{1j}= -\dot{l}_{j}(\bbeta^{(s - 1)})$,  $c_{2j}= -\ddot{l}_{jj}(\bbeta^{(s - 1)})$ and
  $ b_j^{(s)} =
  c_{2j}\beta_j^{(s - 1)} - c_{1j}$\;
  \eIf{$|b_j^{(s)}| < 2 \sqrt{c_{2j} \lambda_n}$}{
   $\beta_j^{(s)} = 0$\;
   }{
   $\beta_j^{(s)} =  \frac{b_j^{(s)} + sign(b_j^{(s)}) \sqrt{ (b_j^{(s)})^2 - 4c_{2j} \lambda_n}}{2c_{2j}}$\;
  }
  }
   \If{$\left\|\bbeta^{(s)} - \bbeta^{(s-1)}\right\| < tol$}{
  $\hat{\bbeta}_{BAR} = \bbeta^{(s)}$ and break\;
   }
  }
\end{algorithm}

\begin{remark}
\label{remark2.4}
(\textsc{cycBAR} versus \textsc{BAR})
The \textsc{cycBAR} algorithm is derived by approximating the log-psuedo likelihood with a quadratic approximation,
so it provides an approximation of the \textsc{BAR} estimator. Because the quadratic approximation is updated iteratively in the algorithm, the difference between them are expected to be mostly negligible, which has been corroborated  by our empirical studies. 
\end{remark}

\begin{remark}
\label{remark2.3}
(Convergence of \textsc{cycBAR})
The \textsc{cycBAR} algorithm resembles the well-known cyclic coordinate decent (CCD) algorithm that has been commonly used for some popular single-step penalized regression methods such as LASSO. However, its numerical convergence is guaranteed by a different mechanism since the
\textsc{cycBAR} algorithm makes coordinate-wise updates for a fixed-point problem whereas CCD aims to decrease an objective function with each coordinate update. Some graphical illustrations of the convergence of the \textsc{cycBAR} algorithm  for $p_n = 2$ are given in Section S2.2 Figures S1 and S2 of the Online Supplementary Material. A rigorous proof of the numerical convergence of the \textsc{cycBAR} algorithm is however not trivial and needs to be investigated in future research. 

\end{remark}


\subsection{Scalable parameter estimation via forward-backward scan}
\label{s2:linear}
Before proceeding further, we note that for the Cox proportional hazards model with no competing risks, $R_i = \{y: X_y \geq X_i\}$ and $\tilde{w}_{ik} \equiv 1$ for all $i$ and $k$. Therefore the score function can be written as
\begin{align}
\label{s2:cox_score}
\dot{l}_j(\bbeta) = \sum_{i=1}^n I(\delta_i = 1) z_{ij} - \sum_{i = 1}^n I(\delta_i = 1) \frac{ \sum_{k \in R_i} z_{kj} \exp(\eta_k)}{\sum_{k \in R_i} \exp(\eta_k)}, 
\end{align}
for $j = 1, \ldots, p_n$. Again, if done directly, calculating $\dot{l}_j(\bbeta)$ will require $O(n^2)$ calculations.  \cite{suchard2013massive} and \cite{mittal2013high}, among others, have implemented the following technique to calculate (\ref{s2:cox_score}) in $O(n)$ calculations. Note that if the event times are arranged in decreasing order, both $\sum_{k \in R_i} z_{kj} \exp(\eta_k)$ and $\sum_{k \in R_i} \exp(\eta_k)$ are a series of cumulative sums. For example, given $X_i > X_{i'}$, the set $R_{i'}$  consists of the observations from $R_i$ and  the set of observations $\{y: X_y \in [X_{i'}, X_{i})\}$, therefore $\sum_{k \in R_{i'}} z_{kj} \exp(\eta_k) = \sum_{k \in R_i} z_{kj} \exp(\eta_k) + \sum_{k \in \{y: X_y \in [X_{i'}, X_{i})\}} z_{kj} \exp(\eta_k)$ and calculating both $\sum_{k \in R_i} z_{kj} \exp(\eta_k)$ and $\sum_{k \in R_i} \exp(\eta_k)$, and consequently its ratio, for all $i = 1, \ldots, n$ will only require $O(n)$ calculations in total. Furthermore, the outer summation of subjects who observe the event of interest is also a cumulative sum since, provided that $X_i > X_{i'}$ and both $\delta_i = 1$ and $\delta_{i'} = 1$,
\begin{align}
\sum_{l = 1}^{i} I(\delta_{l} = 1) \frac{ \sum_{k \in R_l} z_{kj} \exp(\eta_k)}{\sum_{k \in R_l} \exp(\eta_k)} & = \sum_{l = 1}^{i'} I(\delta_{l} = 1) \frac{ \sum_{k \in R_l} z_{kj} \exp(\eta_k)}{\sum_{k \in R_l} \exp(\eta_k)} \\
&  + I(\delta_{i} = 1) \frac{ \sum_{k \in R_{i}} z_{kj} \exp(\eta_k)}{\sum_{k \in R_{i}} \exp(\eta_k)} , 
\end{align}
which will also only require $O(n)$ calculations since the ratio can be precomputed in $O(n)$ calculations. The diagonal elements of the Hessian also follow a similar derivation and can be calculated in $O(n)$ calculations. 

For the PSH model, however, $\sum_{k \in R_i} \tilde{w}_{ik} \exp \left(\eta_j \right)$,
$i=1,\ldots,n$, are not a series of simple cumulative sums because 1) the risk sets $R_i$ are not monotone over time,
and 2)
for each $i$, a different set of weights $\tilde{w}_{ik} = \hat{G}(X_i) / \hat{G}(X_i \mmin X_k),$ $k \in R_i$ are required.
To overcome this problem, we show in Lemma \ref{lem2:factor} below that $\sum_{k \in R_i} \tilde{w}_{ik} \exp \left(\eta_j \right)$ can be decomposed into 
a forward cumulative sum and a backward cumulative sum over two disjoint monotone sets. 
A simple proof is provided in Section S1.6 of the Online Supplementary Material.
\begin{lemma}
\label{lem2:factor}
Assume that no ties are present. Then
\begin{align}
\label{eq2:twosums}
\sum_{k \in R_i} \tilde{w}_{ik} \exp \left(\eta_k \right) 
= \sum_{k \in R_i(1)}   \exp \left(\eta_k \right) +  \hat{G}(X_i) \sum_{k \in R_i(2)}  \exp \left(\eta_k \right) / \hat{G}(X_k)
\end{align}
where $R_i(1) = \{y: (X_y \geq X_i)\}$ and $R_i(2) = \{y: (X_y < X_i \cap \ep_y = 2)\}$ are distinct partitions of $R_i$.
Furthermore, $R_i(1)$ is monotonically decreasing over time and $R_i(2)$ is monotonically increasing over time.
\end{lemma}

Because $R_i(1)$ grows cumulatively as the event times decrease from the largest to the smallest, whereas $R_i(2)$ grows cumulatively as the observed event times increase from the smallest to the largest since it only involves subjects who observed a competing risk and had an observed event time smaller than subject $i$. Thus, similar to the Cox model, the ratio of summations for the score and diagonal Hessian values can be calculated in linear time via a forward-backward scan
where one scan goes in one direction to calculate the cumulative sums associated with $R_i(1)$ and the other scan goes
in the opposite direction to calculate the cumulative sum associated with $R_i(2)$. Therefore, we can effectively reduce the number of operations  from $O(n^2)$ to $O(n)$.

\section{Simulation study}
\label{s2:simulation}

\subsection{Simulation setup}
We simulate datasets under various sample sizes and parameter dimensions. The design matrix, $\mathbf{Z}$
 was generated from a $p_n$-dimensional standard normal distribution with mean zero and pairwise correlation $\mbox{corr}(z_i, z_j) = \rho^{|i-j|}$, where $\rho = 0.5$ simulates moderate correlation. The vector of regression parameters for cause 1, the cause of interest, is $\bbeta_1 = (0.40, 0.45, 0, 0.50, 0, 0.60, 0.75, 0, 0, 0.80, \mathbf{0}_{p - 10})$. 
The data generation scheme follows a similar design to that of \cite{fine1999proportional} and \cite{fu2017penalized}. The CIF for cause 1 is 
$F_1(t; \mathbf{z}_i) = \Pr(T_i \leq t, \epsilon_i = 1|\mathbf{z}_i) = 1 - [1 - \pi\{1-\exp(-t)\}]^{\exp(\mathbf{z}_i^\prime\bbeta_1)}$,
which is a unit exponential mixture with mass $1 - \pi$ at $\infty$ when $\mathbf{z}_i = \mathbf{0}$. Unless otherwise noted, the value of $\pi$ is set to 0.5, which corresponds to a cause 1 event rate of approximately $41\%$. The CIF for cause 2 is obtained by setting 
$\Pr(\epsilon_i = 2 | \mathbf{z}_i) = 1 - \Pr(\epsilon_i = 1|\mathbf{z}_i)$ and then using an exponential distribution with rate $\exp(\mathbf{z}_i^\prime\bbeta_2)$ for the conditional CIF $\Pr(T_i \leq t|\epsilon_i = 2, \mathbf{z}_i)$ with $\bbeta_2 = -\bbeta_1$. Censoring times are independently generated from a uniform distribution $U(0, u_{max})$ where $u_{max}$ controls the censoring percentage. The average censoring percentage for our simulations vary between $30-35\%$.  

\subsection{Finite-sample properties of BAR}
In this section, we briefly summarize the results for comparing the operating characteristics of BAR along with LASSO \citep{tibshirani1996regression}, SCAD \citep{fan2001variable}, 
adaptive LASSO \citep[ALASSO]{zou2006adaptive}, and MCP \citep{zhang2010nearly} which are implemented in the \pkg{crrp} package \citep{fu2017penalized}. Our simulations illustrate that 1)  the BAR estimator is insensitive over the choice of $\xi_n$ over a large interval and 2) BAR performs as well as other oracle-based procedures in terms of estimation and variable selection. This has been observed consistently over several different combinations of model dimension, event rates, signal values, sample sizes, and model sparsity. We refer readers to Section S3 of the Online Supplementary Material for a more detailed explanation of the conclusions from the study.

\subsection{Computational savings via cycBAR and forward-backward scan}
\label{s2:timing}
 In this simulation we illustrate the impressive computational savings obtained from \textsc{cycBAR} and the forward-backward scan described in Sections \ref{s2:cyclic} and \ref{s2:linear}. We compare three implementations of BAR for the PSH model: 
the original BAR without the forward-backward scan,  \textsc{cycBAR}  without the forward-backward scan, and
\textsc{cycBAR}  with the forward-backward scan.
 We let $n$ vary from 600 to 2000, $p_n = 100$, and $\rho = 0.5$ and compute the runtime of each method averaged over 100 simulations.  We report the runtime on a system with an Intel Core i5 2.9 GHz processor and 16GB of memory.
\begin{figure}[h!]
\centering
\includegraphics[scale = 0.58]{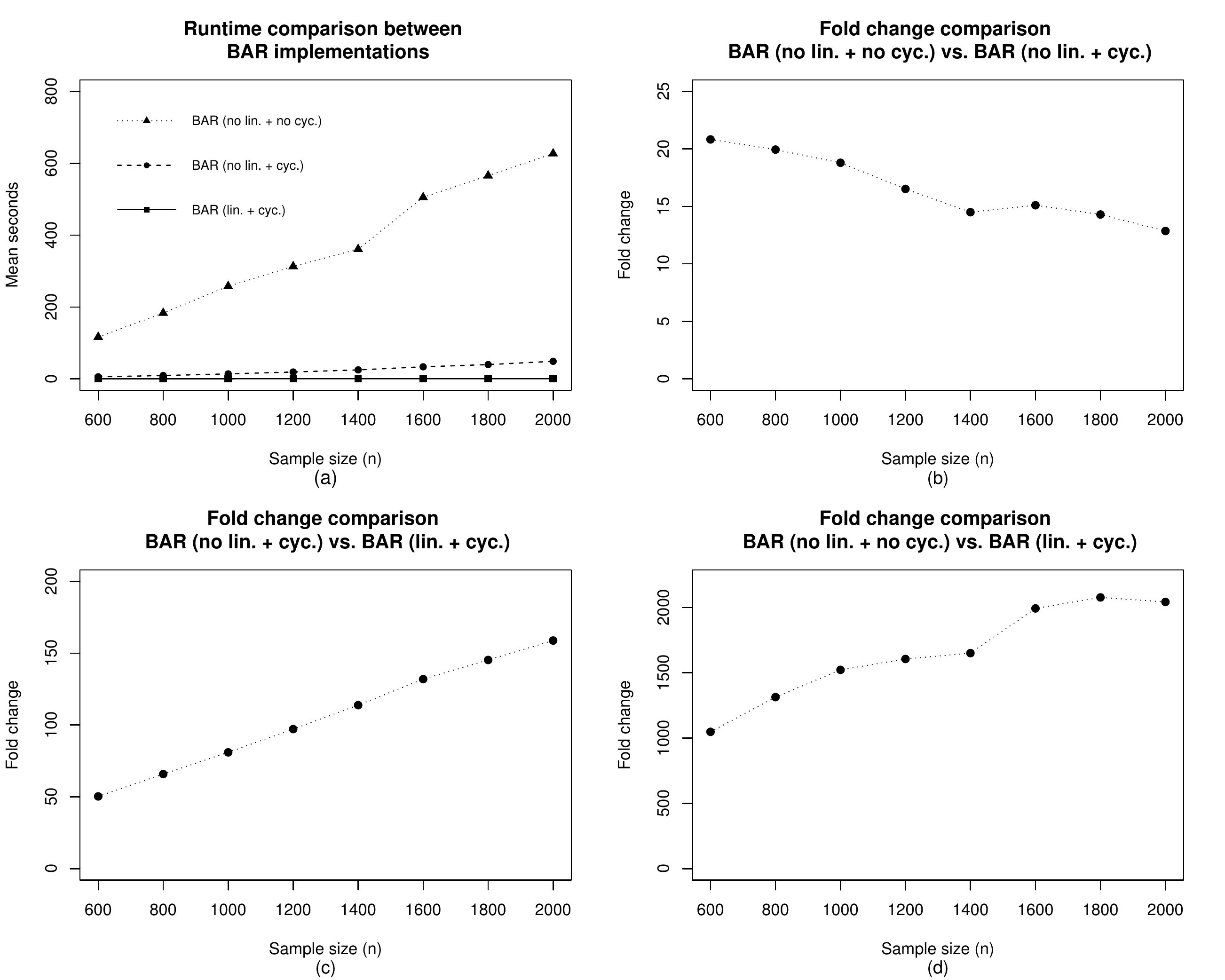}
\label{fig2:timing}
\caption{Runtime comparison between three BAR$(\lambda_n)$ implementations (cyc. = \textsc{cycBAR}  described in Section \ref{s2:cyclic}; lin. = forward-backward scan described in Section \ref{s2:linear}).}
\end{figure}

Figure 1(a) displays the mean runtime (in seconds) for each method as the sample size increases, which shows that
the runtime of the original BAR increases quickly while the runtime of 
 BAR implementing both \textsc{cycBAR} and forward-backward scan
grows at a much slower rate. Panels (b) and (c) further demonstrate the separate contributions of \textsc{cycBAR} and the forward-backward scan method, respectively, using fold change.  Panel (b) shows a 15-20 fold decrease in runtime between \textsc{cycBAR} and the original BAR.  
Panel (c) shows the benefit of linearized estimation, with a 50-150 fold decrease in runtime between \textsc{cycBAR} with and without the forward-backward scan. Additionally, we perform both SCAD and MCP penalizations both with and without the forward-backward scan implementation and observe similar fold changes as observed in Figure 1(c) and the results are tabulated in Table S4 of the Online Supplementary Material. Panel (d) illustrates that using both \textsc{cycBAR} and the forward-backward scan results in a multiplicative gain, yielding an impressive 1,000-2,000 fold speedup in runtime. The runtime reduction is expected to increase as $n$ and/or $p_n$ grow larger as illustrated by the
real data example in the following section.

\section{End-stage renal disease}
\label{s2:data}
The United States Renal Data System (USRDS) is a national data system 
that collects information about end-stage renal disease in the United States.  Patients with end-stage renal disease are known to have a shorter life expectancy compared to their disease-free peers (USRDS Annual Report 2017) and kidney transplantation has been shown to provide better health outcomes for patients with end-stage renal disease \citep{wolfe1999comparison, purnell2016reduced}. 
As an illustration of the scalability of various methods for large data, we run penalized regressions for a PSH model
with 63 demographic and clinical variables using a subset of $n = 225,000$ patients from the USRDS that spans a 10-year study time between January 2005 to June 2015.
 The event of interest was first kidney transplant for patients who were currently on dialysis. Death, renal function recovery, and discontinuation of dialysis are competing risks.  Subjects who are lost to follow up or had no event by the end of study period are considered as right censored. 
 We randomly split the data into a training set ($n = 125,000$) and test set ($n = 100,000$). 
 Table S5 in the Online Supplementary Material shows that the proportions of each type of event are similar across the training and test sets.

The BAR method along with SCAD and MCP penalizations are used to fit the PSH model using the training set.  As with Section \ref{s2:timing}, we consider four implementations of BAR: 1) without both \textsc{cycBAR} and the forward-backward scan;  2) without \textsc{cycBAR} and with the forward-backward scan;
3) with \textsc{cycBAR}  and without the forward-backward scan; and
4) with both  \textsc{cycBAR}  and the forward-backward scan. BIC score minimization, implemented with a 25-value grid search, is used to find the optimal value for the tuning parameter for all three methods. We fix $\xi_n = \log(p_n)$ for the BAR method. SCAD and MCP were performed using the \pkg{crrp} \textsf{R} package \citep{fu2017penalized} where its  
generalized cross validation estimation component is removed to allow a fair comparison of their runtime with BAR only for parameter estimation. Additionally, we run SCAD and MCP penalizations using our forward-backward scan to compare the computational performance of our new implementation to the current state of the art. The BIC score based on the training data is used to compare selection performance between models and predictive performance is measured by the concordance index (c-index) proposed by \cite{wolbers2009prognostic} based on the data. Table \ref{tab2:usrds_results} summarizes the computational time (in seconds), the BIC score,
the c-index, and  the number of selected variables  for each method.  

We observe from Table \ref{tab2:usrds_results} that \textsc{cycBAR}, without the forward-backward scan, took 46 hours to finish, a marked reduction in runtime over the
original BAR implementation which did not finish after 96 hours and was terminated. More impressively, adding the forward-backward scan resulted in an enormous boost in 
speeding up the computation, performing the same task in 40 seconds. We observe similar trends in both SCAD and MCP implementations as well. Our forward-backward scan algorithm results in significant reduction in runtime, over-thousand fold, for BAR, SCAD, and MCP, allowing us to perform variable selection for large-scale competing risks data within seconds rather than days.

The predictive and selection performances of all methods are comparable with similar BIC scores, c-index values and model sizes (number of selected variables), that we attribute to the massive sample size of both the training and test set. The variables selection by BAR are also a subset of the variables selected by both SCAD and MCP. Many of the selected variables 
have been previously reported to have an impact on kidney transplantation.

\begin{table}[t]
\centering
\setlength{\tabcolsep}{5pt}
\caption{Analysis results of a USRDS data using BAR and \textsc{cycBAR} along with MCP and SCAD. (BAR/\textsc{cycBAR}: $\xi_n = \log(p_n)$ and $\lambda_n$ selected through a grid search; BIC was used to select tuning parameters for all methods; Seconds: Runtime in seconds without the forward-backward scan (no scan) and with (scan); BIC score: BIC score based on the training data; $c$-index: $c$-index based on the test data; Model size: Number of nonzero parameters; *The original BAR without \textsc{cycBAR} and  forward-backward scan did not finish
after 96 hours.) 
}
\label{tab2:usrds_results}
\begin{tabular}{r|rrrr}
\hline
  \hline
 &  BAR& \textsc{cycBAR} & SCAD & MCP \\
  \hline
  Seconds (no scan)& 345,600$+^*$& 167,020 & 92,571 & 102,565 \\
  Seconds (scan)&  1,401 & 40 & 37 & 35 \\
    BIC score&  251873.7 & 251867.6 & 251929.9 & 251895.3 \\
  $c$-index&  0.85 & 0.85 & 0.85 & 0.85 \\
  Model size&  43 & 42 & 48 &  49 \\ 
   \hline
\end{tabular}
\end{table}

\section{Discussion}
\label{s2:discussion}
In extending  the $\ell_0$-based BAR methodology to the 
\cite{fine1999proportional} PSH model for competing risks data,  we have developed a novel coordinate-wise update (\textsc{cycBAR}) algorithm to avoid carrying out multiple ridge regressions in the original BAR implementation. Furthermore, we introduce a forward-backward scan algorithm to reduce the computational cost of the log-pseudo likelihood and its derivatives for the PSH model from the order of $O(n^2)$ to $O(n)$. While showing comparable selection and estimation performance, the BAR method for the PSH model using the two new algorithms can produce greater
than 1,000 fold speedups over some current 
penalization methods for the PSH model in numerical studies. 


An important domain of application of the developed scalable sparse regression method is  large comparative effectiveness and drug safety studies using massive electronic health record (EHR) databases such as the Observational Health Data Sciences and Informatics (OHDSI) program
\citep{hripcsak2015observational} (\textcolor{blue}{\it https://ohdsi.org/}) 
and the U.S.~FDA's Sentinel Initiative (\textcolor{blue}{\it https://www.fda.gov/safety/fdassentinelinitiative/ucm2007250.htm}).
These massive databases typically contain millions to hundreds of millions 
patient records with tens of thousands patient attributes,
which are particularly useful  for drug safety studies of a rare event (such as an
unexpected severe adverse event (SAE)) to protect public health. 

As illustrated by the USRDS data example in Section \ref{s2:data}, while existing methods for the PSH model is likely to grind to a halt, the 
developed scalable BAR method with high-performance algorithms has made it possible to analyze these massive data in real time.
To this end, we point out that for a large data with  millions of patient records on tens of thousands covariates, it may not always be feasible to fit a model when the data is stored in the standard dense format due to the high memory requirement. On the other hand, 
these massive datasets are often sparse with only a small portion of covariates are being nonzeros for a given subject. We are currently working on implementing the developed BAR method for sparse massive sample size 
competing risks data by exploiting the sparsity in the data matrix.

Finally, we emphasize that the developed \textsc{cycBAR} method in Section \ref{s2:cyclic} and the forward-backward scan method of Lemma  \ref{lem2:factor} in Section \ref{s2:linear}  are of interest on their own. The \textsc{cycBAR} method can be applied directly to other models and data settings. It is also straightforward to apply the forward-backward scan method to accelerate other estimation methods  for the PSH model. Using this approach, we are currently developing a stand-alone package for \textsf{R} that includes the unpenalized estimation method of \cite{fine1999proportional} and other popular penalization methods.

\section*{Acknowledgement}
 The manuscript was reviewed and approved for publication by an officer of the National Institute of Diabetes and Digestive and Kidney Diseases. Data reported herein were supplied by the USRDS. Interpretation and reporting of these data are the responsibility of the authors and in no way should be seen as official policy or interpretation of the US government. 
Marc A. Suchard's work is partially supported through the National Institute of Health grant U19 AI 135995. The research of Gang Li was partly supported by National Institute of Health Grants P30 CA-16042, UL1TR000124-02, and P50 CA211015.

\bibliographystyle{Chicago}
\bibliography{pshBAR}

\newpage

\appendix

\newpage
\begin{center}
{\large\bf ONLINE SUPPLEMENTARY MATERIAL \\ for ``Scalable Algorithms for Large Competing
Risks Data"}
\end{center}

\section{Statements and technical proofs of theorems and lemmas}
\subsection{Regularity conditions}
Define
\begin{align*}
S^{(k)}(\bbeta, s) & = \frac{1}{n} \sum_{i=1}^n \hat{w}_i(s) Y_i(s)\mathbf{z}_i^{\otimes k} \exp(\bbeta^\prime \mathbf{z}_i), \hspace{.2in} k = 0, 1, 2, \\
\mathbf{E}(\bbeta, s) & = S^{(1)}(\bbeta, s) / S^{(0)}(\bbeta, s),\\
 \mbox{and} \\
V(\bbeta, s) & = S^{(2)}(\bbeta, s)/ S^{(0)}(\bbeta, s) - \mathbf{E}(\bbeta, s)^{\otimes 2},
\end{align*}
where $\mathbf{x}^{\otimes k} = (1, \mathbf{x}, \mathbf{x}\mathbf{x}^\prime)$ for $k = 0, 1, 2$, respectively. Moreover, with  $N_i(t) = I(T_i \leq t, \ep_i = 1)$ and $Y_i(t) = 1 - N_i(t-)$ define
$M_i(\bbeta, t) = \int_0^t dN_i(u) - \int_0^t Y_i(u)h_{10}(u)\exp(\bbeta^\prime \mathbf{z}_i)du$. Similarly, with defining $N_i^c(t) = I(C_i \leq t)$ and $H^c(t)$ being the cumulative hazard function by treating the censored observations as events, $M_i^c(t) = N_i^c(t) - \int_0^t I(X_i \geq u) dH^c(u)$. Let $||\cdot||_p$ be the $\ell_p$-norm for vectors and the norm induced by the vector $p_n$-norm for matrices.  The following technical conditions will be needed in our derivations for the statistical properties of the pshBAR estimator.

 \begin{description}
 \item[(C1)]  $\int_0^\tau h_{01}(t) dt < \infty$;
 \item[(C2)]  There exists some compact neighborhood, $\mathcal{B}_0$, of the true value $\bbeta_0$ such that for $k = 0, 1, 2$, there exists a scalar, vector, and matrix function $s^{(k)}(\bbeta, t)$ defined on $\mathcal{B}_0 \times [0, \tau]$ such that 
  \begin{align*}
\sup_{t \in [0,\tau], \bbeta \in \mathcal{B}_0} \left\| S^{(k)}(\bbeta, t)  - s^{(k)}(\bbeta, t) \right\|_2 = o_p(1), \quad \mbox{as $n\to \infty$};
\end{align*}
\item[(C3)] Let 
$s^{(1)}(\bbeta, t)  = \partial s^{(0)}(\bbeta, t)/ {\partial \bbeta} $ and
$s^{(2)}(\bbeta, t) = \partial s^{(1)}(\bbeta, t)/{\partial \bbeta}$.
For $k = 0, 1, 2$, the functions $s^{(k)}(\bbeta, t)$ are continuous with respect to $\bbeta \in \mathcal{B}_0$, uniformly in $t \in [0, \tau]$, and are bounded on $\bbeta_0 \times [0 ,\tau]$; furthermore,  
$s^{(0)}(\bbeta, t)$ is bounded away from zero on $\mathcal{B}_0 \times [0, \tau]$; 

\item[(C4)] Let $\mathbf{e}(\bbeta, t) = s^{(1)}(\bbeta, t) / s^{(0)}(\bbeta, t)$,
$v(\bbeta, t) = s^{(2)}(\bbeta, t)/s^{(0)}(\bbeta, t) - \mathbf{e}(\bbeta, t)^{\otimes 2}$, and \\
$\Omega = \int_0^\tau v(\bbeta_0, u)s^{(0)}(\bbeta_0, u)h_{10}(u)du$. There exists some constants $C_2$ and $C_3$  such that
\begin{align*}
0 < C_2 < \eigen_{\min}(\Omega) \leq \eigen_{\max}(\Omega) < C_3 < \infty,
\end{align*}
where for any real diagonalizable matrix $\mathbf{A}$, $\eigen_{\min}(\mathbf{A})$  and $\eigen_{\max}(\mathbf{A})$ represent its smallest and largest eigenvalues, respectively; furthermore, there also exists a matrix $\Gamma$ such that $\left\|n^{-1} \sum_{i=1}^n \var(\mathbf{U}_i) - \Gamma \right\|_2 \to 0$, where
\begin{align*}
\mathbf{U}_i = \int_0^\tau \left\{ \mathbf{z}_{i}(u) - \mathbf{e}(\bbeta_0, u) \right\} w_i(u) dM_i(\bbeta_0, u) + \int_0^\tau \mathbf{q}(u)/\pi(u) dM_i^c(u),
\end{align*}
for
\begin{align*}
w_i(t) &  = I(C_i \geq T_i \mmin t)G(t)/G(X_i \mmin t) \\
\mathbf{q}(u) & = -  \lim_{n \to \infty} \frac{1}{n} \sum_{i=1}^n \int_0^\tau \{\mathbf{z}_i(t) - \mathbf{e}(\bbeta_0, t)\}w_i(t)I(X_i < u\leq t) dM_i(\bbeta_0, t) \\
\pi(u) & = \lim_{n \to \infty}  \frac{1}{n} \sum_{i=1}^n I(X_i \geq u)
\end{align*}
\item[(C5)] There exists a constant $C_6$ such that $\sup_{1 \leq i \leq n} E(U_{ij}^2U_{il}^2) < C_6 < \infty$ for all $1 \leq j, l \leq p$, where $U_{ij}$ is the $j$-th element of $\mathbf{U}_i$ defined in (C4); 
\item[(C6)]  As $n\to\infty$, $p_n^4/n \to 0$, $\lambda_n \to \infty$, $\xi_n \to \infty$, $\xi_nb_n/\sqrt{n}  \to 0,$ ${p/(na_n^2)} \to 0$,  $\lambda_nb_n^3\sqrt{q_n}/\sqrt{n} \to 0$ and $\lambda_n\sqrt{q_n}/(a_n^3\sqrt{n}) \to 0$, where $a_n=\min_{j=1, \ldots, q_n} (|\beta_{0j}|)$ and $b_n =\max_{j = 1, \ldots q_n} (|\beta_{0j}|) $. 
\end{description}

The above conditions  (C1)-(C5) are similar to those proposed by \cite{cai2005variable} and \cite{ahn2018group}. Condition (C1) ensures a finite baseline cumulative hazard. 
Condition (C2) ensures the asymptotic stability of $S^{(k)}(\bbeta, t)$, as required under fixed dimension. Under diverging dimension, it follows from Theorem 1 of \cite{kosorok2007marginal} that under certain regularity conditions,
$ 
\sup_{t \in [0,1], \bbeta \in \mathcal{B}_0} \left\| S^{(k)}(\bbeta, t)  - s^{(k)}(\bbeta, t) \right\|_2 \leq \sqrt{p_n \ln p/n},
$
which implies that (C2) holds if $p_n \ln p /n \to 0$. 
Condition (C3) requires that $\exp(\bbeta^\prime\mathbf{z}_i)$ remain integrable under diverging dimension. This will allow integration and differentiation with respect to $S^{(k)}(\bbeta, t)$ ($k = 0, 1$) to be interchanged in our technical derivations.
Condition (C4) guarantees that the covariance matrix of the score function is positive definite and has uniformly bounded eigenvalues for all $n$ and $\bbeta \in \mathcal{B}_0$. Other authors in the variable selection literature have also required a slightly weaker condition  \citep{fan2004nonconcave, cai2005variable, cho2013model, ni2016variable, ahn2018group}. Condition (C5) is vital in proving the Lindeberg condition under diverging dimension for our proof.  Condition (C6) specifies the divergence or convergence rates for the  model size, the penalty tuning parameters, and the lower and upper bound of the true signal. These technical assumptions are only sufficient conditions for our theoretical derivations and observations from our empirical studies illustrate that our theoretical results may, in fact, hold under weaker conditions. Further, we would like to point out that the conditions in (C6) do not impose any one-to-one relationship in finite scenarios. 
  
 \begin{remark}
 \label{re2:ahn2018}
 \cite{ahn2018group} showed that under Conditions (C1) - (C5) and $p_n^4 / n \to 0$
 \begin{align}
\label{eq2:ahn1}
||\dot{l}(\bbeta_0)||_2 = O_p(\sqrt{np_n})
\end{align}
and
\begin{align}
\label{eq2:uniform_hessian}
n^{-1} \ddot{l}(\bbeta) = \Omega + o_p(1),
\end{align}
in probability, uniformly in $\bbeta \in \mathcal{B}_0$. 
 \end{remark}

  \subsection{Statement and proof of the oracle property}
  Let $\bbeta_1$ and $\bbeta_2$ be the first $q_n$  and remaining $p_n - q_n$ components of $\bbeta$, respectively, and define $\bbeta_0 = \left(\bbeta_{01}^\prime,\bbeta_{02}^\prime\right)^\prime$ as the true values of $\bbeta$ where, without loss of generality, $\bbeta_{01}=(\beta_{01} \ldots,\beta_{0{q_n}})$ is a vector of $q_n$ non-zero values and $\bbeta_{02} = \boldsymbol{0}$ is a $p_n-q_n$ dimensional vector of zeros.  Below we state the asymptotic
properties of the BAR estimator for the PSH model under certain regularity conditions. 

\begin{theorem}[Oracle property]
\label{th2:oracle}
Assume the regularity conditions (C1) - (C6)  in the Section S1.1. 
Let  $\hat{\bbeta}_1$ and $\hat{\bbeta}_2$ be the first $q_n$ and the remaining $p_n-q_n$ components of  the BAR estimator $\hat{\bbeta}$, respectively. 
Then, 
\begin{itemize}
\item[(a)] $\hat{\bbeta}_2 = \mathbf{0}$ with probability tending to one;
\item[(b)] $\sqrt{n}\mathbf{d}_n^\prime\Gamma_{11}^{-1/2}\Omega_{11}(\hat{\bbeta}_1 - \bbeta_{01}) \to N( 0, 1)$, for any $q_n$-dimensional vector $\mathbf{d}_n$ such that $||\mathbf{d}_n||_2 \leq 1$ and where $\Gamma_{11}$ and $\Omega_{11}$ are the first $q_n \times q_n$ submatrices of $\Gamma$ and $\Omega$, respectively, defined in Condition (C4).
\end{itemize}

\end{theorem}
Theorem \ref{th2:oracle}(a) establishes that with large probability, the true zero coefficients will be estimated as zeros by the BAR estimator. Part (b) of the theorem
essentially states that the nonzero component of the BAR estimator is asymptotically normal and equivalent to the weighted ridge estimator of the oracle model. To prove Theorem \ref{th2:oracle} we first establish four lemmas. 

 \begin{lemma}[Consistency of ridge estimator]
\label{lem2:ridge}
Let 
\begin{align*}
\hat{\bbeta}_{ridge} & = \arg \min_{\bbeta} \left\{-2l(\bbeta) + \sum_{j=1}^{p_n} \xi_n \beta_j^2\right\}, \\
\end{align*}
be the PSH ridge estimator defined in Equation (3). Assume that Conditions (C1) - (C6) hold. Then 
\begin{equation}
\label{eq2:ridge_result}
||\hat{\bbeta}_{ridge} - \bbeta_0 ||_2 = O_p\left[ \sqrt{p_n}\{n^{-1/2}(1 + \xi_n b_n / \sqrt{n})\}\right]=O_p(\sqrt{p_n/n}),
\end{equation}
where $b_n$ is an upper bound of the true nonzero  $|\beta_{0j}|$'s defined in Condition (C6).
\end{lemma}

{\bf{Proof.}} Let $\alpha_n = \sqrt{p_n}(n^{-1/2} + \xi_nb_n/n)$ and $\ell (\bbeta)= -2l(\bbeta) + \xi_n \sum_{j=1}^{p_n}\beta_j^2$.  To prove Lemma \ref{lem2:ridge}, it is sufficient to show that for any  $\epsilon > 0$, there exists a large enough constant $K_0$ such that
\begin{equation}
\label{eq2:ridge_cond}
\pr \left\{ \inf_{||\mathbf{u}||_2 = K_0} L(\bbeta_0 + \alpha_n\mathbf{u}) > L(\bbeta_0) \right\} \geq 1 - \epsilon,
\end{equation}
since (\ref{eq2:ridge_cond}) implies that there exists a local minimum, $\hat{\bbeta}_{ridge}$, inside the ball $\{\bbeta_0 + \alpha_n \mathbf{u}: ||\mathbf{u}||_2 \leq K_0\}$ such that $||\hat{\bbeta}_{ridge} - \bbeta_0||_2 = O_p(\alpha_n)$, with probability tending to one.  To prove (\ref{eq2:ridge_cond}), we first note 
\begin{align*}
\frac{1}{n} L(\bbeta_0 + \alpha_n \mathbf{u}) - \frac{1}{n}  L(\bbeta_0) 
& = - \frac{1}{n}  \{2l(\bbeta_0 + \alpha_n \mathbf{u}) -  \frac{1}{n}  2l(\bbeta_{0})\} +  \frac{\xi_n}{n}  \sum_{j = 1}^{p_n} \left\{  (\beta_{0j} + \alpha_n u_j)^2 -   \beta_{0j}^2\right\} \\
& = - \frac{1}{n}  \{2l(\bbeta_0 + \alpha_n \mathbf{u}) -  2l(\bbeta_{0})\} + \frac{\xi_n}{n} \sum_{j = 1}^{p_n} \left(2 \beta_{0j}\alpha_nu_j +  \alpha_n^2u_j^2 \right) \\
& \geq -\frac{1}{n}  \{2l(\bbeta_0 + \alpha_n \mathbf{u}) -  2l(\bbeta_{0})\} + \frac{2\xi_n\alpha_n}{n}  \sum_{j = 1}^{p_n}  \beta_{0j}u_j \\
& = - \frac{1}{n}  \{2l(\bbeta_0 + \alpha_n \mathbf{u}) -  2l(\bbeta_{0})\} + \frac{2\xi_n\alpha_n}{n} \sum_{j = 1}^{q_n}\beta_{0j}u_j \\
& \equiv W_1 + W_2.
\end{align*}
By Taylor expansion, we have
\begin{align*}
W_1 & = - \frac{2}{n} \alpha_n \mathbf{u}^\prime\dot{l}(\bbeta_0) - \frac{1}{n} \alpha_n^2 \mathbf{u}^\prime \ddot{l}(\bbeta^*)\mathbf{u} \\
& = W_{11} + W_{12},
\end{align*}
where $\bbeta^*$ lies between $\bbeta_0$ and $\bbeta_0 + \alpha_n \mathbf{u}$, and $\dot{l}(\bbeta)$ and $\ddot{l}(\bbeta)$ denote the first and second derivatives of $l(\bbeta)$, respectively. By the Cauchy-Schwartz inequality, 
\begin{align*}
W_{11} =  -\frac{2}{n} \alpha_n \mathbf{u}^\prime\dot{l}(\bbeta_0) \leq \frac{2}{n} \alpha_n || \dot{l}(\bbeta_0)||_2 \cdot ||\mathbf{u}||_2 =
 \frac{2}{n} \alpha_n O_p(\sqrt{np_n}) ||\mathbf{u}||_2  \leq O_p(\alpha_n^2)||\mathbf{u}||_2,
\end{align*}
where the second equality is due to (\ref{eq2:ahn1}).
By (\ref{eq2:uniform_hessian}) we have 
\begin{align*}
W_{12} & = -\frac{1}{n} \alpha_n^2\mathbf{u}^\prime\ddot{l}(\bbeta^*)\mathbf{u} =  \alpha_n^2\mathbf{u}^\prime\Omega\mathbf{u}\{1 + o_p(1)\}.
\end{align*}
Since $\eigen_{min}(\Omega) \geq C_2 > 0$ by Condition (C4), $W_{12}$ dominates $W_{11}$ uniformly in $||\mathbf{u}||_2 = K_0$ for a sufficiently large $K_0$. Furthermore
\begin{align*}
W_2 & \leq \frac{2\xi_n\alpha_n}{n} | \bbeta_{01}^\prime \mathbf{u} |   \leq \frac{2\sqrt{q_n}\xi_n\alpha_nb_n}{n} ||\mathbf{u} ||_2 = O_p(\alpha_n^2) ||\mathbf{u}||_2,
\end{align*}
where the last step follows from the fact that $\sqrt{q_n}\xi_nb_n/n < \sqrt{p_n}(n^{-1/2} + \xi_nb_n/n) = \alpha_n$. 
Therefore for a sufficiently large $K_0$, we have that $W_{12}$ dominates $W_{11}$  and $W_2$ uniformly in $||\mathbf{u}||_2 =K_0$. Since $W_{12}$ is positive, (\ref{eq2:ridge_cond}) holds and therefore $||\hat{\bbeta}_{ridge} - \bbeta_0||_2 = O_p(\alpha_n) = O_p \left[\sqrt{p_n}\{n^{-1/2}(1 + \xi_n b_n / \sqrt{n})\}\right] =O_p(\sqrt{p_n/n})$, where the last step follows from condition (C6).  
$\Box$

\begin{remark}
Recall $\bbeta = \left(\bbeta_1^\prime, \bbeta_2^\prime\right)^\prime$ where $\bbeta_1^\prime$ and $\bbeta_2^\prime$ correspond to the first $q_n$ and remaining $p_n-q$ components of $\bbeta$, respectively.
Let 
\begin{equation}
\label{eq2:bar_objective}
Q_n(\btheta \mid \bbeta) = -2 l(\btheta) + \lambda_n \btheta^\prime D(\bbeta)\btheta,
\end{equation}
where $D(\bbeta) = diag(\beta_1^{-2}, \beta_2^{-2}, \ldots, \beta_{q_n}^{-2}, \beta_{q_n+1}^{-2}, \ldots, \beta_{p_n}^{-2})$ and $l(\btheta)$  is the $p_n$-dimensional log-partial likelihood of the reduced model. 
Let $\dot{Q}(\btheta \mid\bbeta)$ and $\ddot{Q}(\btheta \mid\bbeta)$ be the first and second derivatives of $Q(\btheta \mid\bbeta)$ with respective to $\btheta$, respectively. Then 
\begin{equation}
\label{eq2:deriv1}
\dot{Q}(\btheta \mid\bbeta) = -2 \dot{l}(\btheta) + 2\lambda_n D(\bbeta)\btheta,
\end{equation}
\begin{equation}
\label{eq2:deriv2}
\ddot{Q}(\btheta \mid\bbeta) = -2\ddot{l}(\btheta) + 2\lambda_n D(\bbeta).
\end{equation}
\end{remark}

\begin{remark}
Let $\hat{\bbeta}_{ridge, 1}$ and $\hat{\bbeta}_{ridge, 2}$ denote the first $q_n$ and the remaining $p_n - q_n$ components of $\hat{\bbeta}_{ridge}$, respectively. 
Then,  Lemma \ref{lem2:ridge} and condition (C6) imply that for $j = 1, \ldots, q_n$ and sufficiently large $n$, $a_n/2 \leq |\hat{\beta}_{ridge, 1j} | \leq 2b_n$, where $\hat{\beta}_{ridge, 1j}$ is the $j^{th}$ component of $\hat{\bbeta}_{ridge, 1}$ and $|| \hat{\bbeta}_{ridge, 2} ||_2 = O(\sqrt{p_n/n})$.
\end{remark}
\begin{remark}
Recall $\bbeta = \left(\bbeta_1^\prime, \bbeta_2^\prime\right)^\prime$ where $\bbeta_1^\prime$ and $\bbeta_2^\prime$ correspond to the first $q_n$ and remaining $p_n-q$ components of $\bbeta$, respectively.
Let 
\begin{equation}
\label{eq2:bar_objective}
Q_n(\btheta \mid \bbeta) = -2 l(\btheta) + \lambda_n \btheta^\prime D(\bbeta)\btheta,
\end{equation}
where $D(\bbeta) = diag(\beta_1^{-2}, \beta_2^{-2}, \ldots, \beta_{q_n}^{-2}, \beta_{q_n+1}^{-2}, \ldots, \beta_{p_n}^{-2})$ and $l(\btheta)$  is the $p_n$-dimensional log-partial likelihood of the reduced model. 
Let $\dot{Q}(\btheta \mid\bbeta)$ and $\ddot{Q}(\btheta \mid\bbeta)$ be the first and second derivatives of $Q(\btheta \mid\bbeta)$ with respective to $\btheta$, respectively. Then 
\begin{equation}
\label{eq2:deriv1}
\dot{Q}(\btheta \mid\bbeta) = -2 \dot{l}(\btheta) + 2\lambda_n D(\bbeta)\btheta,
\end{equation}
\begin{equation}
\label{eq2:deriv2}
\ddot{Q}(\btheta \mid\bbeta) = -2\ddot{l}(\btheta) + 2\lambda_n D(\bbeta).
\end{equation}
\end{remark}

\begin{lemma}
\label{lem2:map}
Let $M_n = \max\{2/a_n, 2b_n\}$. Define $\mathcal{H}_n \equiv\{\bbeta={ \left(\bbeta_1^\prime, \bbeta_2^\prime\right)^\prime}: |\bbeta_1| = (|\beta_1|, \ldots, |\beta_{q_n}|)^\prime  \in [1/M_n, M_n]^{q_n}, 0<\| \bbeta_2 \|_2 \leq \delta_n\sqrt{p_n/n}, \}$, where $\delta_n$ is a sequence of positive real numbers satisfying $\delta_n\rightarrow \infty$ and $p_n\delta_n^2/\lambda_n\rightarrow 0$.
For any given $\bbeta \in \mathcal{H}_n$, define
\begin{equation}
\label{eq2:bar_objective}
Q_n(\btheta \mid \bbeta) = -2 l(\btheta) + \lambda_n \btheta^\prime D(\bbeta)\btheta,
\end{equation}
where $l(\btheta)$  is the $p_n$-dimensional log-partial likelihood and $D(\bbeta) = diag(\beta_1^{-2}, \ldots, \beta_{p_n}^{-2})$. 
Let  $g({\bbeta}) = \left(g_1(\bbeta)^\prime, g_2(\bbeta)^\prime \right)^\prime$ be a solution to $\dot{Q}(\btheta \mid\bbeta) = \mathbf{0}$, where 
\begin{equation}
\label{eq2:deriv1}
\dot{Q}(\btheta \mid\bbeta) = -2 \dot{l}(\btheta) + 2\lambda_n D(\bbeta)\btheta,
\end{equation}
is the derivative of $Q(\btheta \mid\bbeta)$ with respective to $\btheta$.
 Assume that conditions (C1) - (C6) hold. 
 Then, as $n\to \infty$, with probability tending to 1, 
\begin{itemize}
\item[(a)] 
$\sup_{\bbeta \in \mathcal{H}_n}\frac{\| g_2(\bbeta)\|_2}{\|\bbeta_2\|_2} \le  \frac{1}{K_1},
\quad \mbox{for  some constant $K_1 > 1$}$;
\item[(b)] $\left| g_1(\bbeta) \right| \in [1/M_n, M_n]^{q_n}$. 
\end{itemize}
\end{lemma}

{\bf{Proof.}}  By the first-order Taylor expansion and the definition of $g(\bbeta)$, we have 
\begin{equation}
\label{eq2:taylor_q}
\dot{Q}(\bbeta_0|\bbeta) = \dot{Q}\{g(\bbeta) \mid\bbeta\} + \ddot{Q}(\bbeta^{*} \mid \bbeta)\{\bbeta_0 -  g(\bbeta)\} = \ddot{Q}(\bbeta^{*} \mid \bbeta)\{\bbeta_0 -  g(\bbeta)\} ,
\end{equation}
where $\bbeta_0$ is the true parameter vector, and $\bbeta^{*}$ lies between $\bbeta_0$ and $g(\bbeta)$. Rearranging terms, we have 
\begin{equation}
\label{eq2:taylor_q2}
\ddot{Q}(\bbeta^{*} \mid \bbeta)g(\bbeta) = -\dot{Q}(\bbeta_0|\bbeta) + \ddot{Q}(\bbeta^{*} \mid \bbeta)\bbeta_0,
\end{equation}
which can be rewritten as
\begin{align*}
\left\{ -2\ddot{l}(\bbeta^{*}) + 2\lambda_n D(\bbeta) \right\} g(\bbeta) & = -\left\{ -2\dot{l}({\bbeta_0}) + 2 \lambda_n D(\bbeta){\bbeta_0} \right\} + \left\{-2\ddot{l}(\bbeta^{*}) + 2 \lambda_n D(\bbeta) \right\} \bbeta_0 \\
& = 2\dot{l}(\bbeta _0) - 2\ddot{l}(\bbeta^{*}){\bbeta_0}.
\end{align*}
Write $ H(\bbeta) \equiv -n^{-1} \ddot{l}(\bbeta)$, we have
\begin{equation}
\label{eq2:obj0a}
\left\{ H(\bbeta^{*}) + \frac{\lambda_n}{n}D(\bbeta)\right\} g(\bbeta)  = H(\bbeta^{*})\bbeta_0 +  \frac{1}{n}\dot{l}(\bbeta_0),
\end{equation}

which can be further written as 
\begin{equation}
\label{eq2:obj1}
\{g(\bbeta)-\bbeta_0\} + \frac{\lambda_n}{n} H(\bbeta^{*})^{-1} D(\bbeta)g(\bbeta) = \frac{1}{n} H(\bbeta^{*})^{-1} \dot{l}(\bbeta_0).
\end{equation}
Now we partition $H(\bbeta^{*})^{-1}$  into
\[ H(\bbeta^{*})^{-1} = \left[ \begin{array}{ll}
A & B \\
B^\prime & G
\end{array} 
\right] \] 
and  partition $D(\bbeta)$ into 
\[ D(\bbeta) = \left[ \begin{array}{ll}
D_1(\bbeta_1) & \mathbf{0} \\
\mathbf{0}^\prime & D_2(\bbeta_2)
\end{array} 
\right] \] 
where $D_1(\bbeta_1)=\mbox{diag}(|\beta_1|^{-2},...,|\beta_{q_n}|^{-2})$ and $D_2(\bbeta_2)=\mbox{diag}(|\beta_{q_n+1}|^{-2},...,|\beta_{p_n}|^{-2})$. Then (\ref{eq2:obj1}) can be re-written as
\begin{align}
\label{eqnA.15}
\left( \begin{array}{c}
   g_1(\bbeta) - \bbeta_{01} \\
   g_2(\bbeta)
\end{array} \right)  + 
\frac{\lambda_n}{n}
\left( \begin{array}{l}
A D_1(\bbeta_1) g_1(\bbeta) + BD_2(\bbeta_2)g_2(\bbeta)\\
B^\prime D_1(\bbeta_1) g_1(\bbeta) + GD_2(\bbeta_2)g_2(\bbeta) 
\end{array} \right)
= \frac{1}{n} H(\bbeta^{*})^{-1} \dot{l}(\bbeta_0).
\end{align}
Moreover, it follows from (\ref{eq2:ahn1}), (\ref{eq2:uniform_hessian}), and condition (C5)  that
 \begin{equation}\label{eqnA.16}
 \left\| n^{-1}H(\bbeta^{*})^{-1} \dot{l}(\bbeta_0) \right\|_2 = O_p(\sqrt{p_n/n}).
 \end{equation}
  Therefore,
\begin{equation}
\label{eq2:g2eq2}
\sup_{\bbeta \in \mathcal{H}_n}\left\| g_2(\bbeta) + \frac{\lambda_n}{n}B^\prime D_1(\bbeta_1) g_1(\bbeta) + \frac{\lambda_n}{n}GD_2(\bbeta_2)g_2(\bbeta) 
\right\|_2 = O_p(\sqrt{p_n/n}).
\end{equation}
Furthermore, 
\begin{align*}
\left\| g(\bbeta) - \bbeta_0 \right\|_2 &= \left\| -\left\{H(\bbeta^{*}) +  \frac{\lambda_n}{n} D(\bbeta)\right\}^{-1}
\left\{ \frac{\lambda_n}{n} D(\bbeta) \bbeta_0 - \frac{1}{n}\dot{l}(\bbeta_0) \right\}\right\|_2\\
 & \leq  \left\| \left\{H(\bbeta^{*}) \right\}^{-1} 
\left\{ \frac{\lambda_n}{n} D(\bbeta) \bbeta_0 - \frac{1}{n}\dot{l}(\bbeta_0) \right\} \right\|_2 \\
& \leq \left\| \left\{H(\bbeta^{*}) \right\}^{-1} \right\|_2 \cdot \left\{
 \left\|\frac{\lambda_n}{n} D_1(\bbeta_1) \bbeta_{01}\right\|_2 + 
 \left\|\frac{1}{n}\dot{l}(\bbeta_0) \right\|_2  \right\}\\
& = O_p(1) \left\{ O(n^{-1}\lambda_n M_n^{3} \sqrt{q_n} ) + O_p(\sqrt{p_n/n})  \right\} \\
& = O_p(\sqrt{p_n/n}),
\end{align*}
where the first equality follows from (\ref{eq2:obj0a}) and  the fourth step follows from (\ref{eq2:uniform_hessian}), condition (C3), $\left\| n^{-1}\lambda_nD_1(\bbeta_1) \bbeta_{01} \right\|_2 = 
O(n^{-1}\lambda_n M_n^{3} \sqrt{q_n} ) $, and $\left\| n^{-1} \dot{l}(\bbeta_0) \right\|_2 = O_p(\sqrt{p_n/n})$, and the last step holds  since $n^{-1}\lambda_n M_n^{3}\sqrt{q_n} = o(1/\sqrt{n})$ under condition (C6).
Hence,
\begin{align} 
\label{A.15}
\left\| g(\bbeta) \right\|_2  \leq \left\| \bbeta_0 \right\|_2 +\left\| g(\bbeta) - \bbeta_0 \right\|_2 =   O_p(M_n\sqrt{q_n}).
\end{align}
Also note that $\left\| B  \right\|_2 =O_p(1)$ since
$\left\| B B^\prime \right\|_2    \le \left\|A^2 + BB^\prime \right\|_2 + \left\| A ^2\right\|_2 \le 2\left\|A^2 + BB^\prime \right\|_2
 \leq 2\left\| H(\bbeta^{\ast})^{-2} \right\|_2 =O_p(1)$.
This, combined with (\ref{A.15}), implies that
\begin{equation}
\label{eq2:g2eq2}
\sup_{\bbeta \in \mathcal{H}_n}\left\| \frac{\lambda_n}{n}B^\prime  D_1(\bbeta_1) g_1(\bbeta) \right\|_2   \leq \frac{\lambda_n}{n} \sup_{\bbeta \in \mathcal{H}_n} \left\| B \right\|_2 \left\| D_1({\bbeta_1})\right\|_2  \left\| g_1(\bbeta) \right\|_2   
= O_p\left( \frac{\lambda_nM_n^{3} \sqrt{q_n}}{n}\right)=o(1/\sqrt{n}).
\end{equation}
It then follows that 
\begin{align*}
\sup_{\bbeta \in \mathcal{H}_n}\left\| g_2(\bbeta) + \frac{\lambda_n}{n} GD_2(\bbeta_2)g_2(\bbeta) \right\|_2 & \le O_p(\sqrt{p_n/n}) + o(1/\sqrt{n})
 = O_p(\sqrt{p_n/n}).
\end{align*}
Since $G$ is positive definite and symmetric with probability tending to one, by the spectral decomposition theorem, $G = \sum_{i=1}^{p_n-q_n} r_{2i}\mathbf{u}_{2i}\mathbf{u}_{2i}^\prime$, where $r_{2i}$ and $\mathbf{u}_{2i}$ are the eigenvalues and eigenvectors of $G$, respectively. Now with probability tending to one,
\begin{align}
\label{eq2:m2bound1}
\frac{\lambda_n}{n} \left\| G D_2(\bbeta_2)g_2(\bbeta) \right\|_2 & = \frac{\lambda_n}{n} \left\|  \left( \sum_{i=1}^{p_n-q_n} r_{2i}\mathbf{u}_{2i}\mathbf{u}_{2i}^\prime \right) D_2(\bbeta_2)g_2(\bbeta) \right\|_2 \notag \\
& \geq  \frac{\lambda_n}{n} \left\| C_2  \left( \sum_{i=1}^{p_n-q_n} \mathbf{u}_{2i} \mathbf{u}_{2i}^\prime \right) D_2(\bbeta_2)g_2(\bbeta)  \right\|_2 \notag \\
& \geq C_2 \left\| \frac{\lambda_n}{n} D_2(\bbeta_2) g_2(\bbeta) \right\|_2,
\end{align}
where the first inequality is due to (\ref{eq2:uniform_hessian}) and condition (C4) since we can assume that for all $i = 1, \ldots, p - q$,   $r_{2i} \in (C_2, C_3)$ for some $C_2 < C_3 < \infty$ with probability tending to one.
Therefore with probability tending to one,
\begin{align}
\label{eq2:m2bound4}
C_2 \left\| \frac{\lambda_n}{n} D_2(\bbeta_2)g_2(\bbeta) \right\|_2 - \left\| g_2(\bbeta) \right\|_2 \leq \left\| g_2(\bbeta) + \frac{\lambda_n}{n} GD_2(\bbeta_2)g_2(\bbeta) \right\|_2  \leq \delta_n \sqrt{p_n/n},
\end{align}
where $\delta_n$ diverges to $\infty$.
Let $\mathbf{m}_{g_2(\bbeta)/\bbeta_2} = (g_2(\beta_{q_n+1})/\beta_{q_n+1}, \ldots, g_2(\beta_{p_n})/\beta_{p_n})^\prime$. Because
$||\bbeta_2||_2 \leq \delta_n \sqrt{p_n/n}$, we have
\begin{align}
\label{eq2:m2bound2}
C_2 \left\|\frac{\lambda_n}{n} D_2(\bbeta_2) g_2(\bbeta) \right\|_2  = C_2 \frac{\lambda_n}{n} \left\| D_2(\bbeta_2)^{1/2}\mathbf{m}_{g_2(\bbeta)/\bbeta_2} \right\|_2 
\geq  C_2 \frac{\lambda_n}{n} \frac{\sqrt{n}}{\delta_n \sqrt{p_n}} \left\| \mathbf{m}_{g_2(\bbeta)/\bbeta_2} \right\|_2,
\end{align}
and
\begin{equation}
\label{eq2:m2bound3}
\left\| g_2(\bbeta) \right\|_2 = \left\|  D_2(\bbeta_2)^{-1/2}\mathbf{m}_{g_2(\bbeta)/\bbeta_2} \right\|_2 \leq  \left\|  D_2(\bbeta_2)^{-1/2} \right\|_2 \cdot \left\| \mathbf{m}_{g_2(\bbeta)/\bbeta_2} \right\|_2  \leq \frac{\delta_n \sqrt{p_n}}{\sqrt{n}} \left\| \mathbf{m}_{g_2(\bbeta)/\bbeta_2} \right\|_2.
\end{equation}
 Hence it follows from (\ref{eq2:m2bound4}), (\ref{eq2:m2bound2}), and (\ref{eq2:m2bound3}) that with probability tending to one,
\begin{align*}
C_2 \frac{\lambda_n}{n} \frac{\sqrt{n}}{\delta_n \sqrt{p_n}} \left\| \mathbf{m}_{g_2(\bbeta)/\bbeta_2} \right\|_2 - \frac{\delta_n \sqrt{p_n}}{\sqrt{n}} \left\| \mathbf{m}_{g_2(\bbeta)/\bbeta_2} \right\|_2 \leq \delta_n \sqrt{p_n/n}.
\end{align*}
This implies that with probability tending to one,
\begin{equation}
\label{m2bound}
\left\| \mathbf{m}_{g_2(\bbeta)/\bbeta_2} \right\|_2 \leq \frac{1}{\lambda_n/(C_1p\delta_n^2) - 1 }< \frac{1}{K_1},
\end{equation}
for some constant $K_1 > 1$ provided that $\lambda_n/(p_n\delta_n^2) \to \infty$ as $n \to \infty$. Now from (\ref{m2bound}), we have 
\begin{align}\label{m2bound1}
\left\| g_2(\bbeta) \right\|_2 \leq \left\| \mathbf{m}_{g_2(\bbeta)/\bbeta_2} \right\|_2  \max_{q+1 \leq j \leq  p} |\beta_j| \leq \left\|  \mathbf{m}_{g_2(\bbeta)/\bbeta_2} \right\|_2  \left\| \bbeta_2\right\|_2 \leq \frac{1}{K_1} \left\|\bbeta_2 \right\|_2,
\end{align}
with probability tending to one. 
Thus
\begin{align*}
\pr \left( \sup_{\bbeta \in \mathcal{H}_n} \frac{ \left\| g_2(\bbeta) \right\|_2  }{  \left\| \bbeta_2 \right\|_2 } < \frac{1}{K_1} \right) \to 1 \hspace{.2in} \mbox{as $n \to \infty$}
\end{align*}
and (a) is proved.

To prove part (b),  we first note from (\ref{m2bound1}) that as $n\to\infty$,
$
\pr(\left\| \mathbf{m}_{g_2(\bbeta)/\bbeta_2} \right\|_2 \le  \delta_n\sqrt{p_n/n}) \to 1.
$
Therefore it is sufficient to show that for any $\bbeta \in \mathcal{H}_n$, 
$\left| g_1(\bbeta) \right| \in [1/M_n, M_n]^{q_n}$
 with probability tending to 1.
 By  (\ref{eqnA.15}) and (\ref{eqnA.16}), we have 
\begin{align}
\label{g1eq2}
\sup_{\bbeta \in \mathcal{H}_n}\left\| (g_1(\bbeta) - \bbeta_{01} ) + \frac{\lambda_n}{n}A D_1(\bbeta_1) g_1(\bbeta) + \frac{\lambda_n}{n}BD_2(\bbeta_2)g_2(\bbeta) 
\right\|_2=O_p(\sqrt{p_n/n}).
\end{align}
Similar to (\ref{eq2:g2eq2}), it can be shown that 
\begin{align}\label{g1eq2a}
\sup_{\bbeta \in \mathcal{H}_n}\left\| \frac{\lambda_n}{n}A D_1(\bbeta_1) g_1(\bbeta) \right\|_2 = O_p\left( \frac{\lambda_nM_n^{3}\sqrt{q_n}}{n}\right) = o_p(1/\sqrt{n}),
\end{align}
where the last equality holds trivially under condition (C6).
Furthermore,  with probability tending to one,
\begin{align}\label{g1eq2b}
\sup_{\bbeta \in \mathcal{H}_n}\left\| \frac{\lambda_n}{n}BD_2(\bbeta_2)g_2(\bbeta) 
\right\|_2 \leq \frac{\lambda_n}{n} \sup_{\bbeta \in \mathcal{H}_n} \left\| B \right\|_2 \cdot \left\| D_2(\bbeta_2)g_2(\bbeta) \right\|_2   \leq  \frac{\lambda_n}{n} \sqrt{2K_3} \left( \delta_n \sqrt{\frac{p_n}{n}} \right)^2,
\end{align}
for some $K_3>0$, since $||g_2(\bbeta)|| \le \delta_n \sqrt{p_n/n}$, $||B||_2 =O_p(1)$
and $\left\| D_2(\bbeta_2) \right\|_2 \le \delta_n \sqrt{p_n/n}$. Therefore,  combing (\ref{g1eq2}),  (\ref{g1eq2a}) and (\ref{g1eq2b}) gives
\begin{align*}
\sup_{\bbeta \in \mathcal{H}_n} \left\| g_1(\bbeta) - \bbeta_{01} \right\|_2 \leq \frac{\lambda_n}{n} \sqrt{2K_3} \left( \delta_n \sqrt{\frac{p_n}{n}} \right)^2 + \frac{\delta_n\sqrt{p_n}}{\sqrt{n}},
\end{align*}
with probability tending to one. Because  $\lambda_n/n \to 0$ and $\delta_n \sqrt{p_n/n} =\sqrt{p_n\delta_n^2/\lambda_n}  \sqrt{{\lambda_n}/{n}} \to 0$  as $n \to \infty$, we have   
$\pr(\left| g_1(\bbeta) \right| \in [1/M_n, M_n]^{q_n}) \to 1$. This completes the proof of part (b). $\Box$

\begin{lemma}
\label{lem2:asymptotic}
Let $\bbeta_1$ be the first $q_n$ components of $\bbeta$. Define  $f(\bbeta_1) = \arg \min_{\btheta_1} \{ Q_{n1}(\btheta_1 \mid \bbeta_1) \}$,
where
$
Q_{n1}(\btheta_1 \mid \bbeta_1) = -2 l_{n1}(\btheta_1) + \lambda_n \btheta_1^\prime D_1(\bbeta_1)\btheta_1,
$
is a weighted $\ell_2$-penalized -2 log-pseudo likelihood for the oracle model of model size $q_n$, and
$D_1(\bbeta_1) = diag(\beta_1^{-2}, \beta_2^{-2}, \ldots, \beta_{q_n}^{-2})$. Assume that conditions (C1) - (C6) hold. Then with probability tending to one,
\begin{itemize}
\item[(a)] $f(\bbeta_1)$ is a contraction mapping from $[1/M_n, M_n]^{q_n}$ to itself;
\item[(b)] $\sqrt{n}\mathbf{d}_n^\prime  \Gamma_{11}^{-1/2}\Omega_{11}(\hat{\bbeta}_1^{\circ} - \bbeta_{01}) \to N(0, 1)$, for any $q_n$-dimensional vector $\mathbf{d}_n$ such that $\mathbf{d}_n^\prime \mathbf{d}_n = 1$ and  where $\hat{\bbeta}_1^{\circ}$ is the unique fixed point of $f(\bbeta_1)$ and $\Sigma_{11}$ and $\Omega_{11}$ are the first $q_n \times q_n$ submatrices of $\Sigma$ and $\Omega$, respectively.  
\end{itemize}
\end{lemma}

{\bf{Proof:}} (a) First we show that $f(\cdot)$ is a mapping from $[1/M_n, M_n]^{q_n}$ to itself with probability tending to one. Again through a first order Taylor expansion, we have 
 \begin{equation}
 \label{eq2:scoref1}
 \{f(\bbeta_1) - \bbeta_{01}\} + \frac{\lambda_n}{n} H_{1}(\bbeta_1^{*})^{-1}D_1(\bbeta_1)f(\bbeta_1)
 = \frac{1}{n} H_{1}(\bbeta_1^{*})^{-1} \dot{l}_{1}(\bbeta_{01}),
 \end{equation}
 where $H_{1}(\bbeta_1^{*}) = -n^{-1}\ddot{l}_{1}(\bbeta_1^{*})$ exists and is invertible for $\bbeta_1^{*}$ between $\bbeta_{01}$ and $f({\bbeta_1})$. We have 
\begin{align*}
\label{eq2:f1eq2}
\sup_{|\bbeta_1| \in [1/M_n, M_n]^{q_n}} \left\| f(\bbeta_1) - \bbeta_{01} + \frac{\lambda_n}{n}H_{1}(\bbeta_1^{*})^{-1}D_1(\bbeta_1)f(\bbeta_1) \right\|_2 = O_p(\sqrt{q_n/n}),
 \end{align*}
where the right-hand side follows in the same fashion as (\ref{eq2:g2eq2}). Similar to (\ref{eq2:g2eq2}) we have
\begin{align*}
\sup_{|\bbeta_1| \in [1/M_0, M_0]^{q_n}} \left\| \frac{\lambda_n}{n}H_{1}(\bbeta_1^{*})^{-1}D_1(\bbeta_1)f(\bbeta_1) \right\|_2 & = O_p\left(\frac{\lambda_n M_n^3}{\sqrt{n}} \sqrt{\frac{q_n}{n}} \right) = o_p\left( 1/\sqrt{n} \right).  
\end{align*}

Therefore, with probability tending to one
\begin{equation}
\label{eq2:fbound}
\sup_{|\bbeta_1| \in [1/M_n, M_n]^{q_n}} \left\| f(\bbeta_1) - \bbeta_{01} \right\|_2 \leq \delta_n \sqrt{q_n/n},
\end{equation}
where $\delta_n$ is a sequence of real numbers diverging to $\infty$ and satisfies $\delta_n \sqrt{p_n/n} \to 0$.
As a result, we have
\begin{align*}
\pr( f(\bbeta_1) \in [1/M_n, M_n]^{q_n}) \to 1
\end{align*}
as $n \to \infty$. Hence $f(\cdot)$ is a mapping from the region $[1/M_n, M_n]^{q_n}$ to itself. To prove that $f(\cdot)$ is a contraction mapping,  we  need to further show  that
\begin{equation}
\label{eq2:contractresult}
\sup_{|\bbeta_1| \in [1/M_n, M_n]^{q_n}} \left\| \dot{f}(\bbeta_1) \right\|_2 = o_p(1).
\end{equation}
Since $f(\bbeta_1)$ is a solution to $\dot{Q}_{1}(\btheta_1\mid \bbeta_1)= 0$, we have
\begin{equation}
\label{score2}
-\frac{1}{n}\dot{l}_{1}(f(\bbeta_1)) = -\frac{\lambda_n}{n} D_1(\bbeta_1) f(\bbeta_1).
\end{equation}
Taking the derivative of (\ref{score2}) with respect to $\bbeta_1^\prime$ and rearranging terms, we obtain
\begin{align}\label{A.32a}
\left\{ H_{1}(f(\bbeta_1))  + \frac{\lambda_n}{n} D_1(\bbeta_1) \right\} \dot{f}(\bbeta_1) =  \frac{2\lambda_n}{n} diag\{f_1(\bbeta_1)/\beta_1^3, \ldots, f_{q_n}(\bbeta_1)/\beta_{q_n}^3 \}.
\end{align}
With probability tending to one, we have
\begin{align*}
\sup_{|\bbeta_1| \in [1/M_n, M_n]^{q_n}}  \frac{2 \lambda_n}{n} \left\| diag\{f_1(\bbeta_1)/\beta_1^3, \ldots, f_{q_n}(\bbeta_1)/\beta_{q_n}^3 \} \right\|_2 = O_p \left( \frac{\lambda_nM_n^4}{n} \right) = o_p(1),
\end{align*}
where the last step follows from condition (C6). This, combined with (\ref{A.32a}) implies that
\begin{align}
\label{eq2:gradfpt1}
\sup_{|\bbeta_1| \in [1/M_n, M_n]^{q_n}} \left\|  \left\{ H_{1}(f(\bbeta_1)) + \frac{\lambda_n}{n} D_1(\bbeta_1) \right\} \dot{f}(\bbeta_1)\right\|_2 = o_p(1).
\end{align}
Now, it can be shown that probability tending to one, 
\begin{align*}
\left\| H_{1}(f(\bbeta_1))\dot{f}(\bbeta_1) \right\|_2 & \geq \left\|\dot{f}(\bbeta_1) \right\|_2  \cdot \left\|H_{1}(f(\bbeta_1))^{-1} \right\|_2^{-1} \geq \frac{1}{K_2} \left\| \dot{f}(\bbeta_1) \right\|_2,
\end{align*}
for some $K_2>0$, and that
\begin{align*}
\frac{\lambda_n}{n} \left\| D_1(\bbeta_1) \dot{f}(\bbeta_1) \right\|_2&  \geq \frac{\lambda_n}{n} \left\|\dot{f}(\bbeta_1) \right\|_2  \left\|D_1(\bbeta_1)^{-1}\right\|_2^{-1}
 \geq \frac{\lambda_n}{n} \frac{1}{M_n^2} \left\|\dot{f}(\bbeta_1)\right\|_2. 
\end{align*}
Therefore, combining the above two inequalities with (\ref{A.32a}) and (\ref{eq2:gradfpt1}) gives
\begin{align*}
\left( \frac{1}{K_2} - \frac{\lambda_n}{nM_n^2} \right) \sup_{|\bbeta_1| \in [1/M_n, M_n]^{q_n}} \left\|\dot{f}(\bbeta_1)\right\|_2 
= o_p(1).
\end{align*}
This, together with the fact that $\frac{\lambda_n}{n} \frac{1}{M_n^2}=o(1)$, implies that 
 (\ref{eq2:contractresult}) holds. Therefore, with probability tending to one, $f(\cdot)$ is a contraction mapping and consequently has a unique fixed point, say $\hat{\bbeta}_1^{\circ}$, such that $\hat{\bbeta}_1^{\circ} = f( \hat{\bbeta}_1^{\circ})$.
   
We next prove part (b). By (\ref{eq2:scoref1}) we have
\begin{align*}
f(\bbeta_1) = \left\{ H_{1}(\bbeta_1^*) + \frac{\lambda_n}{n}D_1(\bbeta_1) \right\}^{-1}\left\{ H_{1}(\bbeta_1^*)\bbeta_{01} + \frac{1}{n}\dot{l}_{1}(\bbeta_{01}) \right\}.
\end{align*}

Now,
\begin{align}
\sqrt{n} \mathbf{d}_n^\prime \Gamma_{11}^{-1/2}\Omega_{11}(\hat{\bbeta}_1^{\circ} - \bbeta_{01}) & = \sqrt{n} \mathbf{d}_n^\prime \Gamma_{11}^{-1/2}\Omega_{11}  \left[ \left\{ H_{1}(\bbeta_1^*) + \frac{\lambda_n}{n}D_1(\hat{\bbeta}_1^{\circ}) \right\}^{-1} H_{1}(\bbeta_1^*) - I_{q_n} \right] \bbeta_{01} \notag \\
& + \sqrt{n} \mathbf{d}_n^\prime \Gamma_{11}^{-1/2}\Omega_{11} \left[ \left\{ H_{1}(\bbeta_1^*) + \frac{\lambda_n}{n}D_1(\hat{\bbeta}_1^{\circ}) \right\}^{-1} \frac{1}{n} \dot{l}_{1}(\bbeta_{01}) \right] \notag \\
& = I_1 + I_2. \label{eq2:ni0}
\end{align}
Note that for any two conformable invertible matrices $\Phi$ and $\Psi$, we have
\begin{align*}
(\Phi + \Psi)^{-1} = \Phi^{-1} - \Phi^{-1}\Psi(\Phi + \Psi)^{-1},
\end{align*}
Thus we can rewrite $I_1$ as
\begin{align*}
I_1 & = \sqrt{n} \mathbf{d}_n^\prime \Gamma_{11}^{-1/2}\Omega_{11} \left[ \left\{ H_{1}(\bbeta_1^*) + \frac{\lambda_n}{n}D_1(\hat{\bbeta}_1^{\circ}) \right\}^{-1} H_{1}(\bbeta_1^*) - I_{q_n} \right] \bbeta_{01}  \notag \\
& = -\frac{\lambda_n}{\sqrt{n}} \mathbf{d}_n^\prime \Gamma_{11}^{-1/2}\Omega_{11} H_{1}(\bbeta_1^*)^{-1}D_1(\hat{\bbeta}_1^{\circ}) \left\{ H_{1}(\bbeta_1^*) + \frac{\lambda_n}{n}D_1(\hat{\bbeta}_1^{\circ}) \right\}^{-1} H_{1}(\bbeta_1^*)\bbeta_{01}.
\end{align*}
Moreoever
\begin{align} \notag
\left\| I_1 \right\|_2 & \leq \frac {\lambda_n}{\sqrt{n}} \left\|\Gamma_{11}^{-1/2}\Omega_{11} \right\|_2 \left\| H_{1}(\bbeta_1^*)^{-1} \right\|_2 \left\|D_1(\hat{\bbeta}_1^{\circ})\right\|_2
 \left\| \left\{ H_{1}(\bbeta_1^*) + \frac{\lambda_n}{n}D_1(\hat{\bbeta}_1^{\circ}) \right\}^{-1}\right\|_2 \left\| H_{1}(\bbeta_1^*) \right\|_2 \left\| \bbeta_{01} \right\|_2 \\
 & = \frac{\lambda_n}{\sqrt{n}} \cdot O(1) \cdot O_p(1) \cdot M_n^2 \cdot O_p(1) \cdot O_p(1) \cdot M_n\sqrt{q_n} \notag \\
& = O_p(\lambda_n M_n^3\sqrt{q_n}/\sqrt{n}) = o_p(1), \label{eq2:ni1}
\end{align}
where the first equality follows from (\ref{eq2:uniform_hessian}) and condition (C4), and the last equality is a consequence of condition (C6).
Similarly, we can rewrite $I_2$ as
\begin{align}
I_2 & = \sqrt{n}\mathbf{d}_n^\prime \Gamma_{11}^{-1/2}\Omega_{11} \left[ \left\{ H_{1}(\bbeta_1^*) + \frac{\lambda_n}{n}D_1(\hat{\bbeta}_1^{\circ}) \right\}^{-1} \frac{1}{n} \dot{l}_{1}(\bbeta_{01}) \right]  \notag \\
& =  \mathbf{d}_n^\prime \Gamma_{11}^{-1/2}\Omega_{11}H_{1}(\bbeta_1^*)^{-1}\frac{1}{\sqrt{n}}\dot{l}_{1}(\bbeta_{01})  \notag\\
& - \frac{\lambda_n}{\sqrt{n}} \mathbf{d}_n^\prime \Gamma_{11}^{-1/2}\Omega_{11}H_{1}(\bbeta_1^*)^{-1}D_1(\hat{\bbeta}_1^{\circ}) \left\{ H_{1}(\bbeta_1^*)^{-1} + \frac{\lambda_n}{n}D_1(\hat{\bbeta}_1^{\circ}) \right\}^{-1}\frac{1}{n}\dot{l}_{1}(\bbeta_{01}) \notag \\
& =  \mathbf{d}_n^\prime \Gamma_{11}^{-1/2}\Omega_{11}H_{1}(\bbeta_1^*)^{-1}\frac{1}{\sqrt{n}}\dot{l} _{n1}(\bbeta_{01}) + o_p(1).
\label{eq2:i2}
\end{align}
We now establish the asymptotic normality of $n^{-1/2} \mathbf{d}_n^\prime \Gamma_{11}^{-1/2}\Omega_{11}H_{1}(\bbeta_1^*)^{-1}\dot{l}_{1}(\bbeta_{01})$ which will be derived in a similar manner to the proof of Theorem 2 in \citep{cai2005variable}.
By (\ref{eq2:uniform_hessian}), (\ref{eq2:fbound}), and the continuity of $\Omega$, we can deduce that $H_{1}(\bbeta^*) = \Omega_{11} + o_p(1)$. This implies that
\begin{align}
I_2& =
n^{-1/2} \sum_{i=1}^n  \mathbf{d}_n^\prime \Gamma_{11}^{-1/2}\Omega_{11} H_{1}(\bbeta_1^*)^{-1}\mathbf{U}
_{i1} + o_p(1) \notag \\
& = n^{-1/2} \sum_{i=1}^n  \mathbf{d}_n^\prime \Gamma_{11}^{-1/2} \mathbf{U}
_{i1}  
 + \left\{ n^{-1/2} \sum_{i=1}^n \mathbf{d}_n^\prime \Gamma_{11}^{-1/2}\Omega_{11} \mathbf{U}
_{i1}\right\} o_p(1) + o_p(1) \notag \\
& = I_{21} + I_{22} \cdot o_p(1)  + o_p(1),
 \label{eq2:n0}
\end{align}
where $\mathbf{U}_{i1}$ consists of the first $q_n$ components of $\mathbf{U}_i$. 
Letting $Y_{ni} =  n^{-1/2}  \mathbf{d}_n^\prime \Gamma_{11}^{-1/2} \mathbf{U}
_{i1} $,  then by condition (C5)
 \begin{align*}
s_n^2 = \sum_{i=1}^n \var(Y_{ni}) & = \frac{1}{n} \sum_{i=1}^n  \mathbf{d}_n^\prime \Gamma_{11}^{-1/2} \var(\mathbf{U}
_{i1})\Gamma_{11}^{-1/2}\mathbf{d}_n  \\
& = \mathbf{d}_n^\prime \Gamma_{11}^{-1/2} \left\{ \frac{1}{n} \sum_{i=1}^n  \var(\mathbf{U}
_{i1}) \right\} \Gamma_{11}^{-1/2}\mathbf{d}_n \to 1.
\end{align*}
To prove the asymptotic normality of $I_{21}$, we need to verify the Lindeberg condition: for all $\epsilon > 0$,
\begin{align}
\label{eq2:i21_lindeberg_cond}
\frac{1}{s_n^2} \sum_{i=1}^n E\{Y_{ni}^2 I(|Y_{ni}| \geq \epsilon s_n)\}  \to 0,
\end{align}
as $n \to \infty$. Note that
\begin{align}
\label{eq2:i21_lindeberg2}
\sum_{i=1}^n E(Y_{ni}^4) & = n^{-2} \sum_{i=1}^n E\left[\left\{\mathbf{d}_n^\prime \Gamma_{11}^{-1/2} \mathbf{U}
_{i1}\right\}^4\right] \notag \\
& \leq  n^{-2} \sum_{i=1}^nE\left[ || \mathbf{d}_n||_2^4 \cdot ||\Gamma_{11}^{-1/2}||_2^4 \cdot ||\mathbf{U}
_{i1}||_2^4\right] \notag  \\
& = n^{-2} \eigen_{\max}^2\{\Gamma_{11}^{-1} \}\sum_{i=1}^nE(||\mathbf{U}_{i1}||_2^4) \notag \\
& = n^{-2} \eigen_{\max}^2\{\Gamma_{11}^{-1} \} \sum_{i=1}^n \sum_{j=1}^{p_n} \sum_{k = 1}^{p_n} E(U_{ij}^2U_{ik}^2)  \notag \\
& = O(p^2/n),
\end{align}
where the first inequality is due to Cauchy-Schwarz, the second equality is due to $||\mathbf{d}_n||_2 = 1$ and the last step follows from conditions (C4) and (C5). 
Therefore for any $\epsilon > 0$, 
\begin{align*}
\frac{1}{s_n^2} \sum_{i=1}^n E\left\{Y_{ni}^2 I(|Y_{ni}| > \epsilon s_n) \right\} & \leq \frac{1}{s_n^2} \sum_{i=1}^n \left\{ E(Y_{ni}^4) \right\}^{1/2} \left[ E\left\{I(|Y_{ni}| > \epsilon s_n)\right\}^2 \right]^{1/2} \\
& \leq \frac{1}{s_n^2} \left\{\sum_{i=1}^n E(Y_{ni}^4)\right\}^{1/2} \cdot \left\{\sum_{i=1}^n \pr(|Y_{ni}|>\epsilon s_n)\right\}^{1/2} \\
& \leq \frac{1}{s_n^2} \left\{\sum_{i=1}^n E(Y_{ni}^4)\right\}^{1/2} \cdot \left\{\sum_{i=1}^n \frac{\var(Y_{ni})}{\epsilon^2 s_n^2}\right\}^{1/2} \\
& =  \frac{1}{s_n^2} \left\{O(p^2/n)\right\}^{1/2} \frac{1}{\epsilon}\to 0.
\end{align*}
Thus, (\ref{eq2:i21_lindeberg_cond}) is satisfied and
by the Lindeberg-Feller central limit theorem and Slutsky's theorem
\begin{align}
\label{eq2:n1}
I_{21} = 
s_n \left( \frac{1}{s_n} \sum_{i=1}^n Y_{ni}  \right)  \to N(0, 1).
\end{align}
Similarly one can show that $I_{22} = O_p(1)$ and by Slutsky's theorem,
\begin{align*}
n^{-1/2}  \mathbf{d}_n^\prime \Gamma_{11}^{-1/2}\Omega_{11} H_{1}(\bbeta_1^*)^{-1}\dot{l}_{1}(\bbeta_{01}) 
& =n^{-1/2} \sum_{i=1}^n  \mathbf{d}_n^\prime \Gamma_{11}^{-1/2} \mathbf{U}
_{i1} \notag \\ 
& + \left\{ n^{-1/2} \sum_{i=1}^n \mathbf{d}_n^\prime \Gamma_{11}^{-1/2}\Omega_{11} \mathbf{U}
_{i1}\right\} o_p(1) + o_p(1) \notag \\
& = I_{21} + I_{22} \cdot o_p(1) + o_p(1) \\
& \to N(0, 1).  
\end{align*}
Hence, combining (\ref{eq2:ni0}), (\ref{eq2:ni1}), (\ref{eq2:n0}),  and (\ref{eq2:n1}) 
gives
\begin{align*}
\sqrt{n}\mathbf{d}_n^\prime \Gamma_{11}^{-1/2}\Omega_{11}(\hat{\bbeta}_1^{\circ} - \bbeta_{01}) \to N(0, 1),
\end{align*}
which proves part (b). $\Box$ \\

\subsection{Proof of Theorem S1}
Part (a) of the theorem follows immediately from part (a) of Lemma S3.
Part (b) of the theorem will follow from part (b) Lemma S4 and the following
\begin{equation}
\label{eq2:nzero_est}
\Pr \left(\lim_{k \to \infty} \left\| g_1(\bbeta^{(k)}) - \hat{\bbeta}_1^{\circ} \right\|_2 = 0  \right) \to 1,
\end{equation}
where $\hat{\bbeta}_1^{\circ}$ is the fixed point of $f(\bbeta_1)$ defined in Lemma S4.
Note that $g(\bbeta)$ is a solution to 
\begin{equation}
\label{eq2:scorerewrite}
-\frac{1}{n}D(\bbeta)^{-1}\dot{l}(\btheta) + \frac{1}{n}\lambda_n \btheta = \mathbf{0},
\end{equation}
where $D(\bbeta)^{-1} = diag\{ \beta_1^2, \ldots, \beta_{q_n}^2, \beta_{q_n+1}^2, \ldots, \beta_{p_n}^2\}$.
It is easy to see from (\ref{eq2:scorerewrite})  that
\begin{align*}
\lim_{\bbeta_2 \to 0} g_2(\bbeta) = \mathbf{0}_{p_n - q_n}.
\end{align*}
This, combined with (\ref{eq2:scorerewrite}), implies that for any $\bbeta_1$
\begin{align*}
\lim_{\bbeta_2 \to 0} g_1(\bbeta) = f(\bbeta_1).
\end{align*}
Hence, $g(\cdot)$ is continuous  and thus uniform continuous on the compact set $\bbeta \in \mathcal{H}_n$.
Hence as $k\to \infty$, 
\begin{align}\label{omega}
\omega_k \equiv \sup_{|g_1(\bbeta)| \in [1/M_n, M_n]^{q_n}} \left\|  g_1( \bbeta_1, \hat{\bbeta}_2^{(k)}) -f(\bbeta_1) \right\|_2 \to 0,
\end{align}
with probability tending to one. 
Furthermore,
\begin{align}\label{eq2:b1converge}
 \left\| \hat{\bbeta}_1^{(k+1)} - \hat{\bbeta}_1^{\circ} \right\|_2 & \leq \left\| g_1(\hat{\bbeta}^{(k)}) - f(\hat{\bbeta}_1^{(k)}) \right\|_2 + \left\|  f(\hat{\bbeta}_1^{(k)})  - \hat{\bbeta}_1^{\circ} \right\|_2
 \le \omega_k + \frac{1}{K_4} \left\| \hat{\bbeta}_1^{(k)} - \hat{\bbeta}_1^{\circ} \right\|_2,
\end{align}
for some $K_4 > 1$, where the last inequality follows from (\ref{eq2:contractresult}) and the definition of $\omega_k$.
Denote by $a_k =  \left\| \hat{\bbeta}_1^{(k)} -\hat{\bbeta}_1^{\circ} \right\|_2 $, we can rewrite (\ref{eq2:b1converge}) as
\begin{align*}
a_{k+1} \leq \frac{1}{K_4}a_{k} + \omega_k.
\end{align*}
By (\ref{omega}), for any $\epsilon > 0$, there exists an $N > 0$ such that  $\omega_k < \epsilon$ for all $k > N$. Therefore for $k > N$,  
\begin{align*}
a_{k+1} & \leq \frac{1}{K_4}a_{k} + \omega_k \\
& \leq \frac{a_{k-1}}{K_4^2} + \frac{\omega_{k-1}}{K_4}+\omega_k\\
& \leq \frac{a_1}{K_4^k}+\frac{\omega_1}{K_4^{k-1}}+ \cdots+\frac{\omega_N}{K_2^{k-N}}+ \left(\frac{\omega_{N+1}}{K_4^{k-N-1}}+\cdots +\frac{\omega_{k-1}}{K_4}+\omega_k\right)\\
&\le (a_1+\omega_1+ .. .+\omega_N) \frac{1}{K_4^{k-N}} + \frac{1-(1/K_4)^{k-N}}{1-1/K_4} \epsilon
\to 0, \quad\mbox{as $k\to\infty$},
\end{align*}
with probability tending to one. Therefore, 
\begin{align*}
\Pr \left( \lim_{k \to \infty} \left\| \hat{\bbeta}_1^{(k)} - \hat{\bbeta}_1^{\circ} \right\|_2 = \mathbf{0} \right) = 1
\end{align*}
with probability tending to one, or equivalently
\begin{equation}
\label{eq2:result22}
\Pr(\hat{\bbeta}_1 = \hat{\bbeta}_1^{\circ}) = 1
\end{equation}
with probability tending to one. This proves (\ref{eq2:nzero_est}) and thus complete the proof of the theorem.
$\Box$ 

\subsection{Statement and proof of the grouping property}
 An appealing property of $\ell_2$-penalized regression, which does not hold for $\ell_0$-penalized regression, is its tendency to shrink correlated covariates toward each other. As an $\ell_2$-based procedure, the BAR method also retains this grouping property for highly-correlated covariates while retaining the sparsity property of $\ell_0$.
\begin{theorem}[Grouping property]
\label{th2:group}
Assume that $\mathbf{Z} = (\mathbf{z}_i^\prime, \ldots \mathbf{z}_n^\prime)$ is standardized. That is, for all $j = 1, \ldots, p$,
$ \sum_{i=1}^n z_{ij} = 0,\ \mathbf{z}_{[,j]}^\prime\mathbf{z}_{[,j]}  = n - 1,$
where $\mathbf{z}_{[,j]}$ is the $j$th column of $\mathbf{Z}$. Suppose the regularity conditions (C1) - (C6) hold and let $\hat{\bbeta}$ be the BAR estimator. Then for any $\hat{\beta}_i \neq 0$ and $\hat{\beta}_j \neq 0$,
\begin{equation}
\label{eq2:group_bound}
|\hat{\beta}_i^{-1}-\hat{\beta}_j^{-1}|\leq \frac{1}{\lambda_n} \sqrt{2 \{ (n-1)(1 - r_{ij}) \}} \sqrt{n(1+e_n)^2} ,
\end{equation}
with probability tending to one, where $e_n = \sum_{i=1}^n I(\delta_i = 1)$, and  $r_{ij} = \frac{1}{n-1}\mathbf{z}_{[,i]}^\prime\mathbf{z}_{[,j]}$ is the sample correlation of $\mathbf{z}_{[,i]}$ and $\mathbf{z}_{[,j]}$.
\end{theorem}

We can see that as $r_{ij} \to 1$, the absolute difference between $\hat{\beta}_i$ and $\hat{\beta}_{j}$ approaches $0$ implying that the estimated coefficients of two highly correlated variables will be similar in magnitude. \\

{\bf{Proof:}} Under Conditions (C1) - (C6), by Theorem \ref{th2:oracle} we have that $\hat{\bbeta} = \displaystyle \lim_{k \to \infty} \hat{\bbeta}^{(k)}$, where 
\begin{align*}
\hat{\bbeta}^{(k+1)} = g(\hat{\bbeta}^{(k)}) = \mbox{arg} \min_{\bbeta}  \left\{-2l_n(\bbeta) + \lambda_n \sum_{j =1}^{p_n} \frac{I(\beta_j \neq 0) \beta_j^2}{ \left(\hat{\beta_j}^{(k)}\right)^2} \right\}.
\end{align*}
Note that
\begin{equation}
D(\hat{\bbeta}^{(k)})^{-1}\dot{l}_n(\hat{\bbeta}^{(k+1)}) = \lambda_n \hat{\bbeta}^{(k+1)}.\notag
\end{equation}
Therefore for any $l = i, j$ where $\hat{\beta}_i \neq 0$, $\hat{\beta}_j \neq 0$,
\begin{equation}
\label{eq2:derivi}
\hat{\beta}_l^{(k+1)} = \frac{ (\hat{\beta}_l^{(k)})^2}{\lambda_n} \dot{l}_{nl}(\hat{\bbeta}^{(k+1)}). \notag
\end{equation}
Letting $k \to \infty$, (\ref{eq2:derivi}), we have
\begin{equation}
\label{derivi2}
\hat{\beta}_l^{-1} = \frac{1}{\lambda_n} \dot{l}_{nl}(\hat{\bbeta}). \notag
\end{equation}
Letting $\boldeta = Z\bbeta$ we can rewrite the score function 
\begin{equation}
\zeta(\eta_i) = \frac{\partial}{\partial \eta_i} l_n(\boldeta) = \int_0^\tau \hat{w}_i(s)dN_i(s) +   \int_{0}^\tau \frac{ \hat{w}_i^2(s) Y_i(s) \exp(\hat{\eta}_i)}{\sum_{j=1}^n \hat{w}_j(s) Y_{j}(s)\exp(\hat{\eta}_j)} d\bar{N}(s)  \hspace{.2in} i = 1, \ldots, n. \notag
\end{equation}
Recall that $\hat{w}_i(s)Y_i(s) \in [0, 1]$ for all $i = 1, \ldots n$.
Then
\begin{equation}
\left| \zeta(\hat{\eta}_i) \right|   \leq \left| N_i(1) \right| + \left| \int_{0}^\tau \frac{\hat{w}_i^2(s) Y_i(s) \exp(\hat{\eta}_i)}{\sum_{j=1}^n \hat{w}_j(s) Y_{j}(s)\exp(\hat{\eta}_j)} d\bar{N}(s) \right| \leq 1 + e_n \hspace{.2in} i = 1, \ldots, n, \notag
\end{equation}
 where $e_n = \sum_{i = 1}^n I(\epsilon_i = 1)$. Hence
\begin{equation}
\left\| \zeta(\hat{\boldeta}) \right\|_2 \leq  \left\| \boldsymbol{1} + e_n\boldsymbol{1} \right\|_2 = \sqrt{n(1+e_n)^2}. \notag
\end{equation}
Let $\mathbf{z}_{[,i]}$ denote the $i^{th}$ column of $Z$. Since $Z$ is assumed to be standardized, $\mathbf{z}_{[,i]}^\prime\mathbf{z}_{[,i]} = n - 1$ and $\mathbf{z}_{[,i]}^\prime\mathbf{z}_{[,j]} = (n-1)r_{ij}$, for all $i \neq j$ and where $r_{ij}$ is the sample correlation between $\mathbf{z}_{[,i]}$ and $\mathbf{z}_{[,j]}$. Since
\begin{equation}
\hat{\beta}_i^{-1} = \frac{1}{\lambda_n} \mathbf{z}_{[,i]}^\prime \zeta(\hat{\boldeta})  \notag
\hspace{.1in} \mbox{ and } \hspace{.1in}
\hat{\beta}_j^{-1} = \frac{1}{\lambda_n} \mathbf{z}_{[,j]}^\prime \zeta(\hat{\boldeta}), \notag
\end{equation}
we have   
\begin{align}
\left|\hat{\beta}_i^{-1} - \hat{\beta}_j^{-1} \right| & = \left|  \frac{1}{\lambda_n} \mathbf{z}_{[,i]}^\prime  \zeta(\hat{\boldeta})  - \frac{1}{\lambda_n} \mathbf{z}_{[,j]}^\prime \zeta(\hat{\boldeta}) \right|  \notag \\
& = \left| \frac{1}{\lambda_n} (\mathbf{z}_{[,i]} - \mathbf{z}_{[,j]})^\prime \zeta(\hat{\boldeta}) \right| \notag  \\
& \leq \frac{1}{\lambda_n} \left\| (\mathbf{z}_{[,i]} - \mathbf{z}_{[,j]}) \right\| \left\| \zeta(\hat{\boldeta}) \right\| \notag \\
& \leq \frac{1}{\lambda_n} \sqrt{2\{ (n-1) - (n-1)r_{ij}\}} \sqrt{n(1+e_n)^2}  \notag
\end{align}
for any $\hat{\beta}_i \neq 0$ and $\hat{\beta}_j \neq 0$. $\Box$

\newpage

\subsection{Proof of Theorem 1}
Because $\hat\bbeta$ is a fixed point of $g(\cdot)$ or $\hat\bbeta =g(\hat\bbeta)$, we have
for $j=1, \ldots, p$
\begin{align}
\{\tilde{\mathbf{X}}^\prime\tilde{\mathbf{X}} + \lambda_n D(\hat\bbeta)\} 
\begin{pmatrix}
\label{eq2:ols_decomp}
          \begin{bmatrix}
           \hat\beta_1 \\           
           \vdots \\
           0 \\
           \vdots \\
           \hat\beta_p
          \end{bmatrix} +
          \begin{bmatrix}
           0 \\
           \vdots \\
           \hat\beta_j \\
           \vdots \\
           0
         \end{bmatrix}
    \end{pmatrix}
        = \tilde{\mathbf{X}}^\prime\tilde{\mathbf{y}}.
\end{align}
Alternative, one can rewrite (\ref{eq2:ols_decomp}) as
\begin{align}
\label{eq2:ols_decomp2}
\{D(\hat\bbeta)^{-1}\tilde{\mathbf{X}}^\prime\tilde{\mathbf{X}} + \lambda_n \mathbf{I}_{p}\} 
	\begin{pmatrix}
          \begin{bmatrix}
           \hat\beta_1 \\           
           \vdots \\
           0 \\
           \vdots \\
           \hat\beta_p
          \end{bmatrix} +
          \begin{bmatrix}
           0 \\
           \vdots \\
           \hat\beta_j \\
           \vdots \\
           0
         \end{bmatrix}
    \end{pmatrix}
        = D(\hat\bbeta)^{-1}\tilde{\mathbf{X}}^\prime\tilde{\mathbf{y}}.
\end{align}
By extracting the $j${th} element of (\ref{eq2:ols_decomp2}), we have
\begin{align}
\label{eq2:univar_extract}
\tilde{\mathbf{x}}_j^\prime \sum_{i \neq j} \tilde{\mathbf{x}}_i \hat\beta_i^3 + \lambda_n\cdot 0 + \tilde{\mathbf{x}}_j^\prime\tilde{\mathbf{x}}_j \cdot \hat\beta_j^3 + \lambda_n \hat\beta_j = \tilde{\mathbf{x}}_j^\prime\tilde{\mathbf{y}}\hat{\beta}_j^2,
\end{align}
Letting $b^*_j = \tilde{\mathbf{x}}_j^\prime(\tilde{\mathbf{y}} -  \sum_{i \neq j} \tilde{\mathbf{x}}_i \hat\beta_i)$, simple algebra will allow us to rewrite (\ref{eq2:univar_extract}) as
\begin{align}
\label{eq2:univar_extract2}
\hat\beta_j ( \tilde{\mathbf{x}}_j^\prime\tilde{\mathbf{x}}_j \cdot \hat\beta_j^2 - b^*_j \hat\beta_j 
+ \lambda_n)  = 0,
\end{align}
which yields
\begin{align}
\label{eq2:ccd_quadratic2}
 \hat\beta_j = 
 \begin{cases} 
      0, & \mbox{ if $|b_j^*| < 2\sqrt{\tilde{\mathbf{x}}_j^\prime\tilde{\mathbf{x}}_j  \lambda_n} $}\\
     \frac{b^*_j + sign(b^*_j) \sqrt{ (b^*_j)^2 - 4\tilde{\mathbf{x}}_j^\prime\tilde{\mathbf{x}}_j  \lambda_n}}{2\tilde{\mathbf{x}}_j^\prime\tilde{\mathbf{x}}_j },
 & \mbox{otherwise.}
   \end{cases}
\end{align}
for $j=1,...,p$.
$\qquad\Box$

\subsection{Proof of Lemma 1}
Recall that, for the PSH model, $\tilde{w}_{ik} = \hat{G}(X_i) / \hat{G}(X_k \mmin X_i)$. Because $R_i = \{y:(X_y \geq X_i) \cup (X_y \leq X_i \cap \epsilon_y =  2)\}$, $k \in R_i$ implies that either $k \in \{y:(X_y \geq X_i)\}$ or $k \in \{y:(X_y \leq X_i \cap \epsilon_y =  2)\}$. If $k \in \{y:(X_y \geq X_i)\}$, then $\tilde{w}_{ik} = \hat{G}(X_i) / \hat{G}(X_i) = 1$. If $k \in \{y:(X_y \leq X_i \cap \epsilon_y =  2)\}$, then $\tilde{w}_{ik} = \hat{G}(X_i) / \hat{G}(X_k)$. Therefore
\begin{align*}
\sum_{k \in R_i} \tilde{w}_{ik} \exp \left(\eta_k \right) & = \sum_{k \in R_i(1)} \tilde{w}_{ik} \exp \left(\eta_k \right)  + \sum_{k \in R_i(2)} \tilde{w}_{ik} \exp \left(\eta_y \right) \\
& = \sum_{k \in R_i(1)}   \exp \left(\eta_k \right) +  \hat{G}(X_i) \sum_{k \in R_i(2)}  \exp \left(\eta_k \right) / \hat{G}(X_k), \notag 
\end{align*}
where $R_i(1) = \{y: (X_y \geq X_i)\}$ and $R_i(2) = \{y: (X_y < X_i \cap \ep_y = 2)\}$.

\newpage
\section{Supplementary material for Section 2.3}

\subsection{BAR regression via cyclic coordinate descent}
\RestyleAlgo{boxruled}
\LinesNumbered
\begin{algorithm}[h!]
  \footnotesize
\label{alg2:ccd}
\SetAlgoLined
 Set $\hat{\bbeta}^{(0)} = \hat{\bbeta}_{ridge}$\;
  \For{$k= 1, 2, \ldots$}{
  $\bbeta^{(0)} = \hat{\bbeta}^{(k-1)}$\;
 \For{$s= 1, 2, \ldots$}{
$\#$ Enter cyclic coordinate descent \\
  \For{$j = 1, \ldots p$}{
  Calculate 
  $c_{1j}= \dot{l}_{j}(\bbeta^{(s - 1)})$ and $c_{2j}= -\ddot{l}_{jj}(\bbeta^{(s - 1)})$\;
  $ \bbeta_j^{(s)} = (c_{2j}\beta_j^{(s - 1)} + c_{1j}) / \{c_{2j} + \lambda_n/ (\hat\bbeta_j^{(k-1)})^2\}$\;
   }
     \If{ $\left\| \bbeta^{(s)} - \bbeta^{(s-1)} \right\| < tol_1$ }{
  $\hat\bbeta^{(k)} = \bbeta^{(s)}$ and break\;
   }
  }
 \If{$\left\| \hat\bbeta^{(k)} - \hat\bbeta^{(k-1)} \right\| < tol_2$}{
   $\hat{\bbeta}_{BAR} = \hat\bbeta^{(k)}$ and break \;
   }
  }
  $\hat{\bbeta}_{BAR} = \hat{\bbeta}_{BAR} \times I(|\hat{\bbeta}_{BAR}| > \epsilon^*)$ $\#$ Induce sparsity\;
 \caption{The \textsc{BAR} algorithm using cyclic coordinate descent optimization }
\end{algorithm}

\subsection{Computational behavior of \textsc{cycBAR}}
 We illustrate
under a simple scenario with $p_n=2$ that the \textsc{cycBAR} algorithm converges to the fixed point of $g(\beta_1, \beta_2)$ along the graphs of $\beta_1 = g_1(\beta_2)$ and $\beta_2 = g_2(\beta_1)$, with each coordinate-wise update moving monotonically a step closer to the fixed point. 
\begin{figure}[h]
\centering
\includegraphics[scale = 0.6]{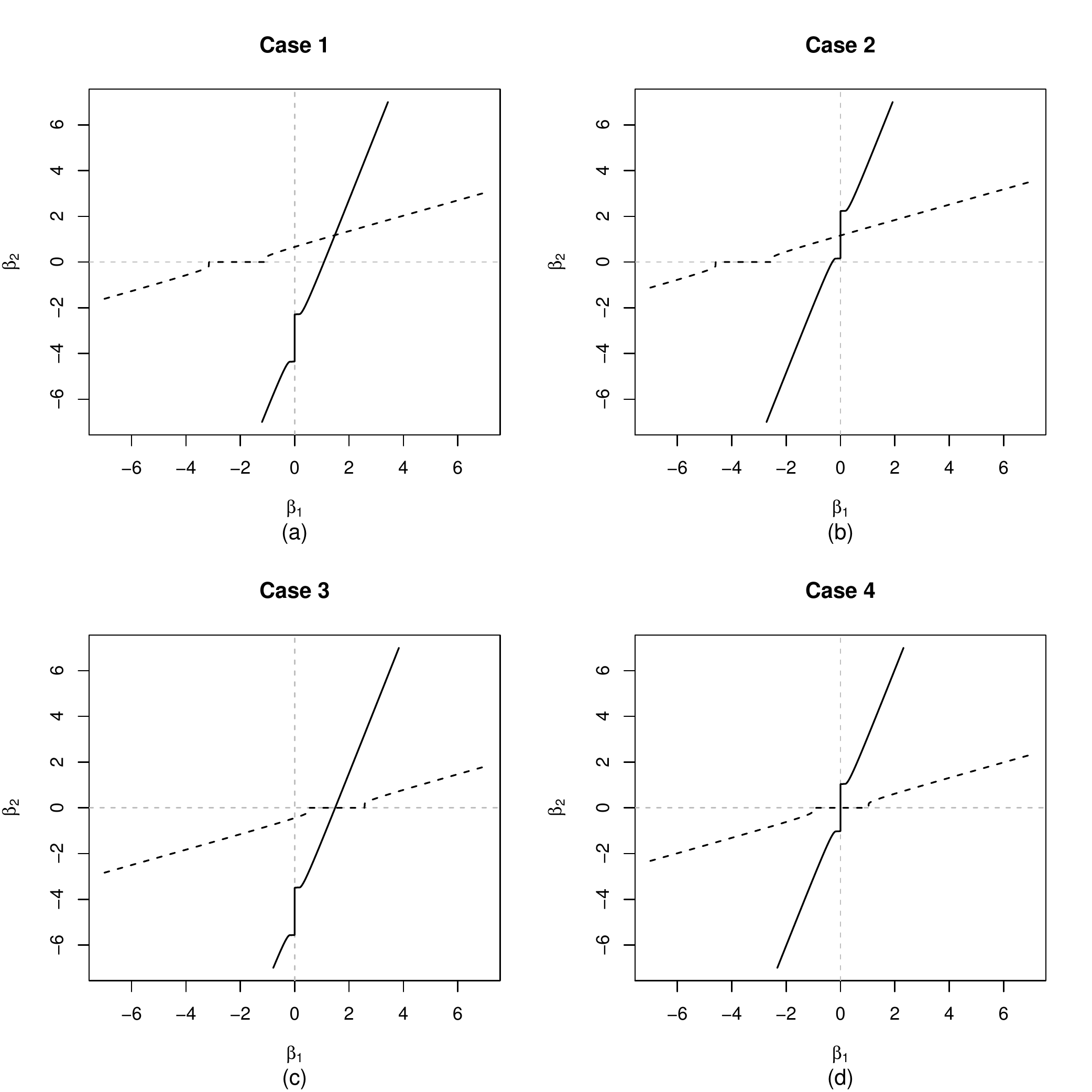}
\caption{Graphs of $\beta_1 = g_1(\beta_2)$ (solid line) and $\beta_2 = g_2(\beta_1)$ (dotted line) under selected scenarios, which by Theorem 1,
intersect at the fixed-point  of $g(\beta_1, \beta_2)$. }
\end{figure}

\newpage
\begin{figure}[h!]
\centering
\includegraphics[scale = 0.6]{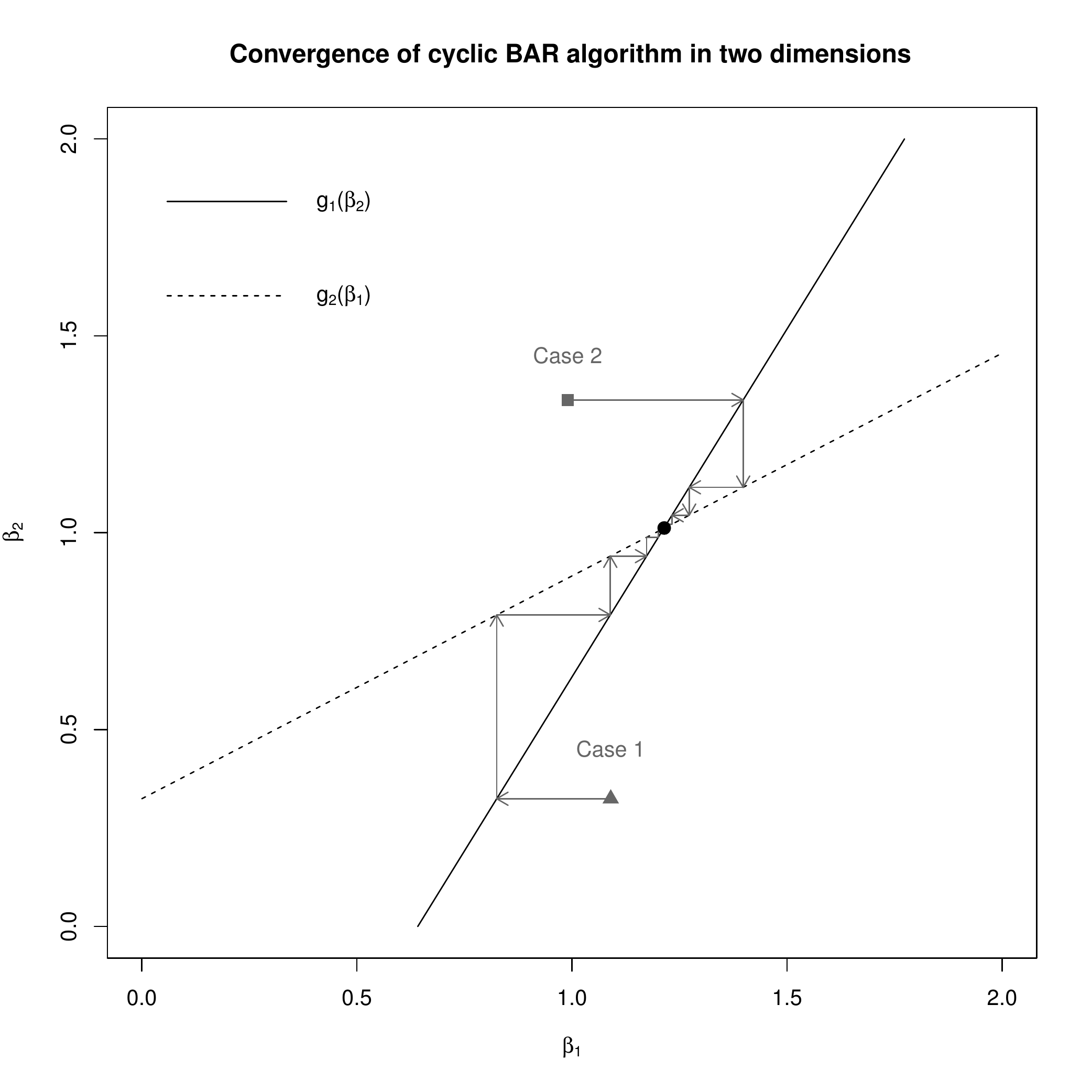}
\label{fig2:cwplot}
\caption{An illustration of the \textsc{cycBAR} algorithm in a zoomed in picture of Figure S1(a). The BAR estimator is the fixed point of $g(\beta_1, \beta_2)$, which, by Theorem 1, is the intersection of $\beta_1 = g_1(\beta_2)$ and $\beta_2 = g_2(\beta_1)$.}
\end{figure}

\section{Supplementary material for Section 3.2}
The operating characteristics of BAR with different tuning parameter selection strategies along with LASSO, adaptive LASSO (ALASSO), SCAD and MCP are assessed by the following measures.  As a gold standard, we also fit the oracle model (ORACLE) as if the true model was known {\em{a priori}}. Estimation bias is summarized through the mean squared bias (MSB), $E\{\sum_{i=1}^{p} (\hat{\bbeta}_i - \bbeta_{0i})^2\}$. Variable selection performance is measured by a number of indices: the mean number of false positives (FP), the mean number of false negatives (FN);  and average similarity measure  (SM) for support recovery where $SM=||\hat{\mathcal{S}}_1 \cap \mathcal{S}_1||_0/\sqrt{||\hat{\mathcal{S}}_1||_0 \cdot ||\mathcal{S}_1||_0}$ and $\mathcal{S}_1$ and $\hat{\mathcal{S}}_1$ are the set of indices for the non-zero components of $\bbeta_1$ and $\hat{\bbeta}_1$, respectively \citep{zhang2017simultaneous}. The similarity measure can be viewed as a continuous measure for true model recovery:  it is close to 1 when the estimated model is similar to the true model, and close to 0 when the estimated model is highly dissimilar to the true model. 

For BAR, we investigate two tuning parameter selection approaches: 1)  $\xi_n$ and $\lambda_n$ are selected via a two-dimensional grid search to minimize the BIC criterion (BAR$(\xi_n, \lambda_n)$); and 2)  $\lambda_n$ is selected via grid search to minimize the BIC criterion; and $\xi_n = \log(p_n)$ (BAR$(\lambda_n)$). The grids for $\xi_n$ and $\lambda_n$ were chosen from a log-spaced interval of 25 values between $[0.001, 3\log(p_n)]$. Unless otherwise noted, we implement BAR using both the \textsc{cycBAR} and forward-backward scan. The tuning parameter for LASSO, ALASSO, SCAD, and MCP is selected by minimizing the BIC-score through a data-driven grid search of 25 possible values for $\lambda_n$. We only consider the $p_n < n$ scenario and thus use the maximum pseudo likelihood estimator as the initial estimator for ALASSO.

Tables S1-S3 display the estimation and selection performances of BAR with LASSO, ALASSO, SCAD, and MCP across several simulation scenarios. The selection and estimation performances between optimizing over both $\xi_n$ and $\lambda_n$ (BAR($\xi_n, \lambda_n$)) and over only $\lambda_n$ (BAR($\lambda_n$)) are similar, suggesting that the BAR estimator is insensitive over the choice of $\xi_n$. This is further corroborated by Figures S3-S5 where the solution path of the BAR estimator with various choices of are stable
over a large interval of $\xi_n$. We also observe that BAR and MCP are generally top performers in every scenario and that, as expected, LASSO tends to select more noise variables. 

\newpage
\begin{table}[h!]
\centering
\setlength{\tabcolsep}{5pt}
\caption{Additional simulation results for model comparison. Based on 100 replications with $\rho = 0.5$, $\bbeta_1 = (\bbeta^*, \mathbf{0}_{p_n - 10})$ where $\bbeta^* = (0.40, 0.45, 0, 0.50, 0, 0.60, 0.75, 0, 0, 0.80)$, censoring rate $\approx 33\%$ and type 1 event rate $\approx 41\%$.}
\begin{tabular}{r| rrrr  | rrrr}
   \hline
    \hline
   &  \multicolumn{4}{c}{$n = 300;p = 100$} &  \multicolumn{4}{c}{$n = 700; p = 100$}  \\
    \cline{1-9}
    Method & MSB & FN & FP & SM &  MSB & FN & FP & SM    \\
    \cline{1-9}
ORACLE & 0.09 & 0.00 & 0.00 & 1.00 & 0.04 & 0.00 & 0.00& 1.00 \\ 
\hline
  BAR$(\xi_n, \lambda_n)$ & 0.31 & 0.40 & 1.87 & 0.85 & 0.06 & 0.01 & 0.89 & 0.94 \\ 
  BAR$(\lambda_n)$ & 0.32 & 0.49 & 1.70  & 0.85 & 0.06 & 0.01 & 0.86 & 0.94 \\  
  LASSO & 0.44 & 0.10 & 2.82 & 0.83 & 0.21 & 0.00 & 2.49 & 0.85 \\ 
  ALASSO & 0.39 & 0.75 & 2.00 & 0.81 & 0.09 & 0.00 & 0.73 & 0.95 \\ 
  SCAD & 0.43 & 0.33 & 2.73 & 0.82 & 0.12 & 0.02 & 1.39 & 0.91 \\ 
  MCP & 0.37 & 0.56 & 1.89 & 0.84 & 0.08 & 0.08 & 0.65 & 0.95 \\  
   \hline
\end{tabular}
\end{table}

\begin{table}[h!]
\centering
\setlength{\tabcolsep}{5pt}
\caption{Additional simulation results for model comparison. Based on 100 replications with $\rho = 0.5$, $\bbeta_1 = (\bbeta^*, \mathbf{0}_{p_n - 10})$ where $\bbeta^* = (0.40, 0.45, 0, 0.50, 0, 0.60, 0.75, 0, 0, 0.80)$, censoring rate $\approx 33\%$ and type 1 event rate $\approx 32\% (\pi = 0.4)$ and $\approx 43\% (\pi = 0.75)$.}
\begin{tabular}{r| rrrr  | rrrr}
   \hline
    \hline
   &  \multicolumn{4}{c}{$n = 700; p = 100; \pi = 0.4$} &  \multicolumn{4}{c}{$n = 700; p = 100;\pi = 0.75$}  \\
    \cline{1-9}
    Method & MSB & FN & FP & SM  &  MSB & FN & FP & SM    \\
    \cline{1-9}
ORACLE & 0.04 & 0.00 & 0.00 & 1.00 & 0.03 & 0.00 & 0.00 &1.00 \\ 
\hline
  BAR$(\xi_n, \lambda_n)$ & 0.08 & 0.03 & 0.88 & 0.94 & 0.04 & 0.00 & 0.65 & 0.96 \\ 
  BAR$(\lambda_n)$ & 0.08 & 0.04 & 0.84 & 0.94 & 0.05 & 0.00 & 0.67 & 0.95  \\ 
  LASSO & 0.23 & 0.00 & 2.63 & 0.85 & 0.18 & 0.00 & 2.51 & 0.86 \\ 
  ALASSO & 0.11 & 0.06 & 0.94 & 0.93 & 0.06 & 0.00 & 0.58 & 0.96 \\ 
  SCAD & 0.15 & 0.08 & 1.44 & 0.90 & 0.09 & 0.00 & 0.96 & 0.93 \\ 
  MCP & 0.11 & 0.13 & 0.73 & 0.94 & 0.06 & 0.02 & 0.35 & 0.97 \\ 
  \hline
\end{tabular}
\end{table}

\newpage
\begin{table}[h!]
\centering
\setlength{\tabcolsep}{5pt}
\caption{Additional simulation results for model comparison. Based on 100 replications with $\rho = 0.5$, $\bbeta_1 = (\bbeta^*, \bbeta^*, \bbeta^*, \mathbf{0}_{p_n - 30})$ where $\bbeta^* = (0.40, 0.45, 0, 0.50, 0, 0.60, 0.75, 0, 0, 0.80)$, censoring rate $\approx 33\%$ and type 1 event rate $\approx 41\%$.}
\begin{tabular}{r| rrrr  | rrrr}
   \hline
    \hline
   &  \multicolumn{4}{c}{$n = 300; p = 100$} &  \multicolumn{4}{c}{$n = 700; p = 100$}  \\
    \cline{1-9}
    Method & MSB & FN & FP & SM  &  MSB & FN & FP & SM    \\
    \cline{1-9}
ORACLE & 0.40 & 0.00 & 0.00  & 1.00 & 0.13 & 0.00 & 0.00 & 1.00 \\ 
\hline
  BAR$(\xi_n, \lambda_n)$ & 0.84 & 0.24 & 3.60 & 0.91 & 0.16 & 0.01 & 1.38  & 0.96 \\ 
  BAR$(\lambda_n)$ & 0.79 & 0.36 & 3.27 & 0.91 & 0.16 & 0.01 & 1.33 & 0.97 \\ 
  LASSO & 2.32 & 0.05 & 11.02 & 0.79 & 1.27 & 0.00 & 11.92 & 0.78 \\
  ALASSO & 1.21 & 0.57 & 6.35 & 0.85 & 0.32 & 0.00 & 2.40  & 0.94 \\
  SCAD & 0.98 & 0.19 & 7.00 & 0.85 & 0.16 & 0.01 & 1.54 & 0.96 \\  
  MCP & 1.03 & 0.33 & 3.59  & 0.91 & 0.15 & 0.02 & 0.73 & 0.98 \\ 
   \hline
\end{tabular}
\end{table}

\newpage
\begin{figure}[h]
\centering
\includegraphics[scale = 0.75]{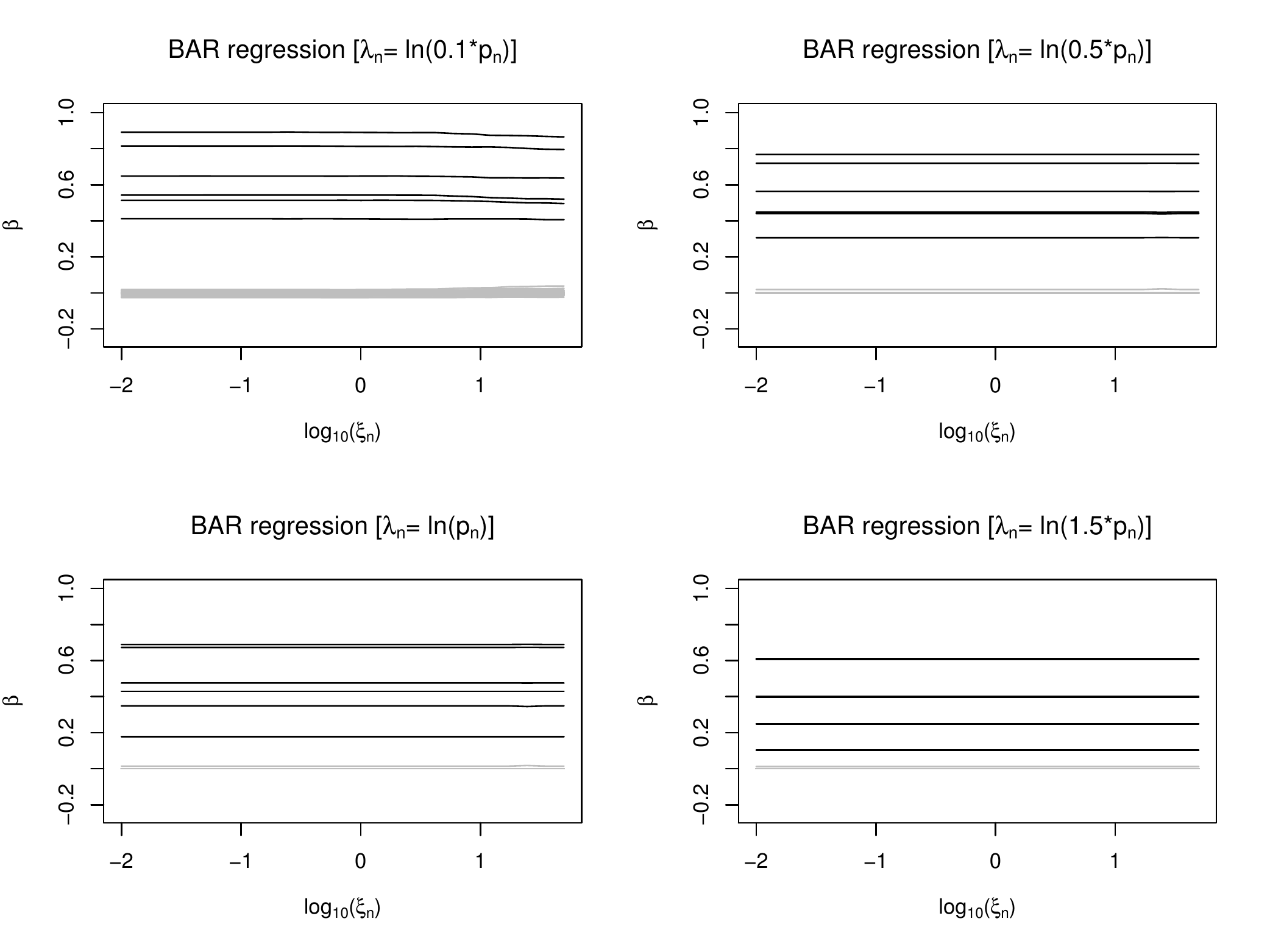}
\caption{Path plot for BAR regression with varying $\xi_n$ and several fixed values of $\lambda_n$ where $n = 300$ and $p_n = 40$. The path plots are averaged over 100 simulations.}
\end{figure}

\newpage
\begin{figure}[H]
\centering
\includegraphics[scale = 0.75]{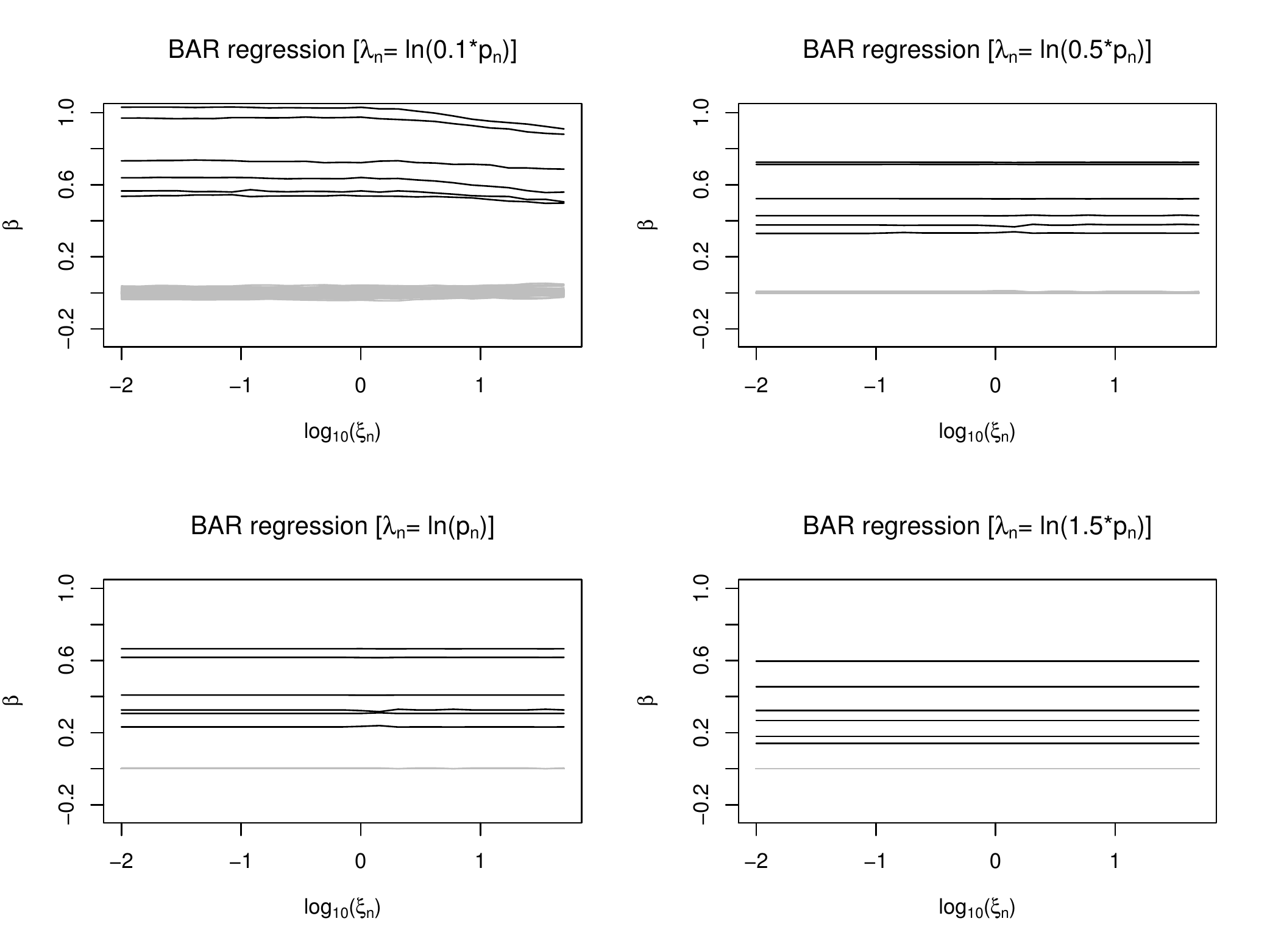}
\caption{Path plot for BAR regression with varying $\xi_n$ and several fixed values of $\lambda_n$ where $n = 300$ and $p_n = 100$. The path plots are averaged over 100 simulations.}
\end{figure}

\newpage
\begin{figure}[h!]
\centering
\includegraphics[scale = 0.75]{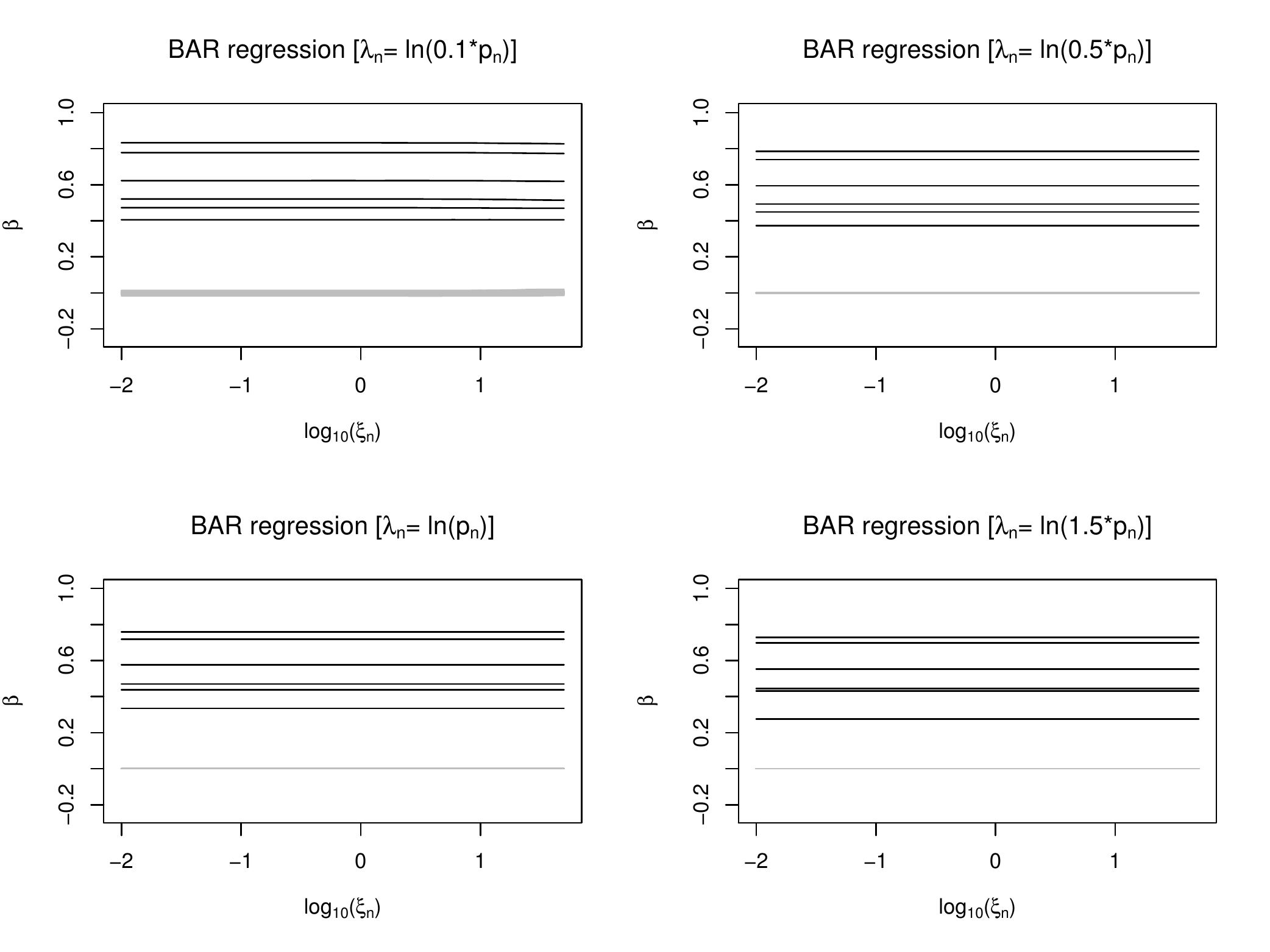}
\caption{Path plot for BAR regression with varying $\xi_n$ and several fixed values of $\lambda_n$ where $n = 700$ and $p_n = 40$. The path plots are averaged over 100 simulations.}
\end{figure}

\section{Supplementary material for Section 3.3}
\begin{table}[h!]
\centering
\caption{SCAD and MCP penalizations performed with (scan) and without (no scan) the forward-backward scan. Runtime is calculated as mean seconds over 100 simulations.}
\setlength{\tabcolsep}{5pt}
\begin{tabular}{lr|rrrrrrrr}
\hline
\hline
 & & $n=$ 600 & 800 & 1000 & 1200 & 1400 & 1600 & 1800 & 2000 \\ 
  \hline
SCAD & (scan) & 0.16 & 0.17 & 0.20 & 0.23 & 0.28 & 0.30 & 0.31 & 0.33 \\ 
  & (no scan) & 4.82 & 7.77 & 12.29 & 15.11 & 23.65 & 35.26 & 34.96 & 42.91 \\ 
  \hline
  MCP& (scan) & 0.21 & 0.23 & 0.27 & 0.27 & 0.32 & 0.38 & 0.39 & 0.42 \\ 
  & (no scan) & 6.17 & 9.06 & 14.99 & 15.49 & 24.56 & 34.07 & 32.65 & 46.67 \\ 
   \hline
\end{tabular}
\end{table}
  
  \begin{table}[h!]
\centering
\caption{Additional simulation results for model comparison. Based on 100 replications with $\rho = 0.5$, $\bbeta_1 = (\bbeta^*, \mathbf{0}_{p_n - 10})$ where $\bbeta^* = (0.40, 0.45, 0, 0.50, 0, 0.60, 0.75, 0, 0, 0.80)$, censoring rate $\approx 33\%$ and type 1 event rate $\approx 41\%$.}
\label{tab2:add_1}
\setlength{\tabcolsep}{5pt}
\fbox{
\begin{tabular}{r| rrrr  | rrrr}
   &  \multicolumn{4}{c}{$n = 300;p = 100$} &  \multicolumn{4}{c}{$n = 700; p = 100$}  \\
    \cline{1-9}
    Method & MSB & FN & FP & SM &  MSB & FN & FP & SM    \\
    \cline{1-9}
ORACLE & 0.09 & 0.00 & 0.00 & 1.00 & 0.04 & 0.00 & 0.00& 1.00 \\ 
\hline
  BAR$(\xi_n, \lambda_n)$ & 0.31 & 0.40 & 1.87 & 0.85 & 0.06 & 0.01 & 0.89 & 0.94 \\ 
  BAR$(\lambda_n)$ & 0.32 & 0.49 & 1.70  & 0.85 & 0.06 & 0.01 & 0.86 & 0.94 \\  
  BAR$_{EBIC}$ & 0.51 & 1.68 & 0.03 & 0.84 & 0.10 & 0.25 & 0.00 & 0.98 \\ 
  LASSO & 0.44 & 0.10 & 2.82 & 0.83 & 0.21 & 0.00 & 2.49 & 0.85 \\ 
  ALASSO & 0.39 & 0.75 & 2.00 & 0.81 & 0.09 & 0.00 & 0.73 & 0.95 \\ 
  SCAD & 0.43 & 0.33 & 2.73 & 0.82 & 0.12 & 0.02 & 1.39 & 0.91 \\ 
  MCP & 0.37 & 0.56 & 1.89 & 0.84 & 0.08 & 0.08 & 0.65 & 0.95 \\  
\end{tabular}
}
\end{table}

\section{Supplementary material for Section 4}
\begin{table}[H]
\centering
\label{tab3:prop}
\caption{Additional information about the USRDS subset used in Section 4. Summary of event count (\%) observed for the training ($n = 125,000$) and test ($n = 100,000$) sets for the USRDS subset. (Disc: Discontinued dialysis; Recov: Renal function recovery; RC: Right censored including loss-to-follow up and end of study time.)}
\setlength{\tabcolsep}{3.5pt}
\begin{tabular}{r|rrrrr|r}
  \hline
  \hline
Set & Transplant & Death & Disc.& Recov. & RC & Total\\ 
 \hline
  Training &  11,943($10\%$) & 60,175 ($48\%$) & 8,160 ($6\%$) & 7,555 ($6\%$) & 37,167 ($30\%$) & 125,000 ($100\%$)\\
  Test & 9,642 ($10\%$) & 47,830 ($48\%$) & 6,459 ($7\%$) & 6,057 ($6\%$) & 30,012 ($29\%$) & 100,000 ($100\%$)\\
   \hline
\end{tabular}
\end{table}

\pagenumbering{arabic}

\numberwithin{equation}{section}
\numberwithin{figure}{section}
\numberwithin{table}{section}
\end{document}